\documentclass[showpacs,showkeys,preprintnumbers,pre,amsmath,amssymb,superscriptaddress]{revtex4}
\pdfoutput=1

\usepackage{graphicx}
\usepackage{hyperref}
\usepackage{graphics}
\usepackage{epsfig}
\usepackage{bm}
\usepackage{epic,eepic}
\usepackage{epsfig}
\usepackage{pifont}

\usepackage{caption}
\usepackage{caption}
\usepackage{subfigure}

\usepackage{xspace}
\usepackage{graphicx,color}
\usepackage{amsmath,amssymb,amsfonts,mathrsfs}
\usepackage{url}
\usepackage[normalem]{ulem}
\usepackage{booktabs}
\usepackage{float}

\usepackage[toc,page]{appendix}
\usepackage[capitalize]{cleveref}
\usepackage{mathtools}
\usepackage[normalem]{ulem}

\providecommand{\st}[1]{{}^{_{\text{#1}}}\!}

\newcommand{\be}{\begin{equation}}
\newcommand{\ee}{\end{equation}}
\newcommand{\ben}{\begin{equation*}}
\newcommand{\een}{\end{equation*}}

\newcommand{\bv}[1]{\mathbf{#1}}

\newcommand{\vv}{\bv{v}}

\newcommand{\delF}[1]{\textcolor{blue}{}}

\newcommand{\tcap}{t_{\mbox{\tiny cap}}}
\newcommand{\ellT}{\ell_{\mbox{\tiny T}}}

\begin{document}

\title{Effects of thermal fluctuations in the fragmentation of a nano-ligament}

\author{X. Xue}
\email{X.Xue@tue.nl}
\affiliation{Department of Physics  and J.M. Burgerscentrum, Eindhoven University of Technology, 5600 MB Eindhoven, the Netherlands.}
\affiliation{Department of Physics \& INFN, University of Rome ``Tor Vergata'', Via della Ricerca Scientifica 1, 00133, Rome, Italy.}

\author{M. Sbragaglia}
\email{sbragaglia@roma2.infn.it}
\affiliation{Department of Physics \& INFN, University of Rome ``Tor Vergata'', Via della Ricerca Scientifica 1, 00133, Rome, Italy.}

\author{L. Biferale}
\email{biferale@roma2.infn.it}
\affiliation{Department of Physics \& INFN, University of Rome ``Tor Vergata'', Via della Ricerca Scientifica 1, 00133, Rome, Italy.}

\author{F. Toschi}
\email{F.Toschi@tue.nl}
\affiliation{Departments of Physics and of Mathematics and Computer Science and J.M. Burgerscentrum, Eindhoven University of Technology, 5600 MB Eindhoven, the Netherlands.}
\affiliation{ Istituto per le Applicazioni del Calcolo CNR, Via dei Taurini 19, 00185 Rome, Italy.}

\begin{abstract}
We study the effects of thermally induced capillary waves in the fragmentation of a liquid ligament into multiple nano-droplets. Our numerical implementation is based on a fluctuating lattice Boltzmann (LB) model for non-ideal multicomponent fluids, including non-equilibrium stochastic fluxes mimicking the effects of molecular forces at the nanoscales. We quantitatively analyze the statistical distribution of the break-up times and the droplet volumes after the fragmentation process, at changing the two relevant length scales of the problem, i.e., the thermal length-scale and the ligament size. The robustness of the observed findings is also corroborated by quantitative comparisons with the predictions of sharp interface hydrodynamics. Beyond the practical importance of our findings for nanofluidic engineering devices, our study also explores a novel application of LB in the realm of nanofluidic phenomena. The post print version has been published in: ``Phys. Rev. E 98, 012802 DOI:10.1103/PhysRevE.98.012802"
\end{abstract}

\maketitle

\section{Introduction}\label{sec:intro}
The hydrodynamical description of a non-ideal interface at the microscales is generically assessed via the combined effects of viscous dissipation and surface tension forces~\cite{Christopher07,Seeman12,Christopher08,Teh08,Baroud10,RaucherReview08,Bocquet10}. Pushing such description towards smaller scales faces two main difficulties. From one side, hydrodynamics itself cannot hold true at all scales of motion, and its coarse-grained foundations are inevitably weakened whenever the physics gets closer to atomistic level. From the other side, if hydrodynamics needs to be corrected by atomistic effects, one faces the issue of the appropriate description to adopt. The equations of fluctuating hydrodynamics~\cite{Landau, Zarate2006} provide a promising route, accounting for molecular collisions via the introduction of stochastic contributions to the non-equilibrium fluxes, as originally proposed by Landau~\cite{Landau}. Such treatment has the major appeal to retain the generic hydrodynamic approach, since it does not enter into a detailed characterization of the atomistic motion, but rather takes a coarse-grained perspective, where molecular effects are modeled via the fluctuation-dissipation balance~\cite{kubo1966fluctuation}. So far, a consistent body of work considered various aspects of interfacial flows in open and confined geometries~\cite{Baroud10,Seeman12,Craster09,Oron97,Eggers97,EggersVillermaux08}, while there are only very few works considering the effect of thermal fluctuations~\cite{MoroStone05,Moro15,Grun06,Petit12,Moseler00,Eggers02,Gross14,Hennequin06}. At small scales, thermal fluctuations promote interface excitations with energy $k_B T$; these are resisted by the surface tension $\gamma$, which opposes a force (per unit length) against the deformation of the interface. This balance determines a new length scale, named {\it thermal lengthscale}, defined as ~\cite{shi1994cascade} 
$$
\ell_{\mbox{\tiny T}}=\sqrt{\frac{k_BT}{\gamma}},
$$ 
which is typically in the nanometer range (or fractions of it)~\cite{Grant,Safran}. On general grounds, thermal fluctuations are expected to become increasingly more relevant and produce measurable effects when moving from micrometer scales down to smaller nanometer scales. This is the case, for example, of nanojets, were stochastic hydrodynamic equations have been used to study the interface dynamics~\cite{Moseler00,Eggers02} with extensive comparisons with fully atomistic descriptions~\cite{Moseler00}. It was convincingly shown that thermal fluctuations impact the break-up properties of nanojets. For the spreading of viscous drops on a solid substrate, thermal fluctuations have been shown to accelerate the spreading in comparison to the deterministic case~\cite{tanner1979spreading}. This has been studied with the help of fluctuating hydrodynamics in the lubrication limit~\cite{MoroStone05}, and later confirmed by mesoscale numerical simulations~\cite{Gross14}. Other numerical methods were used to study dewetting of thin liquid films, and it was observed that thermal fluctuations accelerate the rupture of films~\cite{Grun06}.
Also, experimental studies exist~\cite{Hennequin06,Petit12}, concerning the break-up of a nanojet in the presence of thermal fluctuations. In these studies, it is observed that thermal noise suppresses the formation of satellite droplets~\cite{Hennequin06}, but no quantitative characterization of the distribution of droplets size has been provided. In the study~\cite{Petit12}, the pinch-off process has also been found to be affected by thermal fluctuations, since an initial visco-capillary regime is followed by a fluctuation-dominated regime when the characteristic size of the neck approaches the thermal length.

The focus of this paper is on the quantitative analysis of the effects of thermal fluctuations on the statistics of the break-up times and on the droplet volume distribution, following the contraction and fragmentation of a liquid nano-ligament. Similarly to other studies~\cite{Moseler00,Eggers02} we rely on a hydrodynamical description coupled to the effects of thermal noise. However, we do not solve the continuum equations of interfacial hydrodynamics with a sharp interface directly, but instead, utilize the fluctuating multicomponent lattice Boltzmann (LB)~\cite{Belardinelli15}. The study here presented contributes to show a realistic application of fluctuating LB for nanofluidic phenomena. Our main findings can be summarized as follows: we confirm that thermal fluctuations are able to accelerate nano-ligament fragmentations process, as it was previously reported~\cite{Moseler00,Eggers02}. Furthermore, we give quantitative information on how much the thermal noise can accelerate the break-up process, and we present effects of thermal fluctuations on the polydispersity of droplets distribution. Last but not least, we find that the LB simulations with thermal noise are consistent with sharp interface hydrodynamics results. The paper is organized as follows: in \cref{sec:immisciblephases} we briefly review the fluctuating LB used; in \cref{sec:ligamentsetup} we present technical details of the numerical simulations; in \cref{sec:PRI} we report on the destabilization process driven by the Plateau-Rayleigh instability and the qualitative effects of thermal fluctuations; in \cref{sec:breakup} we discuss results on the statistics of break-up times, while in \cref{sec:PDFV} we report on the statistics of the droplet volumes; in \cref{sec:hydroLBM} we provide quantitative comparison between the LB results and the results of sharp interface hydrodynamics; conclusions will follow in \cref{sec:conclusions}.


\section{Model: Fluctuating lattice Boltzmann for multi-component fluids}\label{sec:immisciblephases}
 
Beyond the traditional problems of homogeneous hydrodynamics~\cite{Benzi92,Aidun10,ChenDoolen98}, LB models have proven particularly suitable for the modeling of complex fluids with multiple phases and/or components~\cite{Zhang11}. Moreover, stimulated by earlier contributions for homogeneous fluids~\cite{Ladd,Adhikari2005,Dunweg07}, recently there has been a significant work to include the effects of thermal fluctuations in LB for multiphase~\cite{Gross10, Varnik11} and multicomponent flows~\cite{Thampi11,Belardinelli15}. Technical details of the fluctuating LB have already been extensively presented in~\cite{Belardinelli15}, and here we only briefly recall the most important facts for the sake of completeness. We employ the D3Q19 LB model, which discretizes the momentum lattice into 19 directions. The method describes the physics of a mixture with 2 fluid components (say $A$ and $B$) in terms of probability distributions functions $f_{l i}(\mathbf{x},t)$ evaluated at a lattice position $\mathbf{x}$ at time $t$, with $i$ being a discrete index associated to a discrete velocity, $\mathbf{c}_{i} (i = 0,...,18)$, and $l$ being the index for the fluid component ($l = A, B$). The distribution function is updated via the combined effect of streaming, collisions, interaction forces and stochastic noise:
\be
\label{eq:lbe}
f_{l i}(\mathbf{x}+\mathbf{c}_{i},t+1) -f_{l i}(\mathbf{x}, t)=\mathfrak{L}(f_{l i}(\mathbf{x},t )) + F_{l i}(\mathbf{x},t) + \xi _{l i}(\mathbf{x},t)  \hspace{.2in} l=A,B\\
\ee
where $\mathfrak{L}$ is a collision kernel, $F_{l i}$ is a source coming from non-ideal forces, and $\xi _{li}$ is a stochastic source. For simplicity we have used a unitary time step. The macroscopic quantities such as density $\rho_{l}$ (one for each component), and global velocity $\vv$ are readily evaluated from the distribution functions: 
\begin{equation}\label{eq:density}
\rho_{l}(\mathbf{x}, t) = \sum_{i} f_{l i}(\mathbf{x}, t),  \hspace{.2in} \vv(\mathbf{x}, t) = \frac{\sum_{i,l} f_{l i}(\mathbf{x}, t)\mathbf{c}_{i}}{\rho_{\st{tot}}(\mathbf{x}, t)}
\end{equation}
where $\rho_{\st{tot}}=\rho_A+\rho_B$ is the total density. Regarding the collisional operator $\mathfrak{L}$, we use a MRT (multi-relaxation time) scheme~\cite{DHumieres02,Dunweg07,SchillerThesis}. The basic idea behind the MRT scheme is to introduce a vector basis $\mathbf{e}_{n} (n = 0,...,18)$ to decompose the probability distribution functions into ``modes'',  $M_{l n} = \Sigma_{i} \mathbf{e}_{n i} f_{l i}$. The lowest order modes coincide with hydrodynamic modes (density, momentum, stress tensor) while higher order modes (``ghost'' modes) do not contribute to the hydrodynamic behaviour of the LB models~\cite{Dunweg07, DHumieres02}. Each one of the modes is relaxed with its own relaxation frequency towards the corresponding equilibrium mode calculated from the equilibrium distribution
\be
f_{l i}^{(\mbox{\tiny eq})}(\mathbf{x}, t) = \rho_{l}(\mathbf{x}, t)\, \omega_{i} \left(1+\frac{\mathbf{c}_{i} \cdot \vv (\mathbf{x}, t)}{c_{s}^{2}} + \frac{(\mathbf{c}_{i} \cdot \vv(\mathbf{x}, t))^{2}}{2c_{s}^{4}} - \frac{\vv^{2}(\mathbf{x}, t)}{2c_{s}^{2}} \right)
\ee
where $c_{s}$ is the speed of sound (a constant in the model) and $\omega_{i}$ are weights associated to the discrete lattice directions. The non-ideal forces are chosen in the Shan-Chen formulation~\cite{SC93,SC94,Zhang11,SbragagliaBelardinelli,SegaSbragaglia13}:
\begin{equation}\label{sc-force_eq}
F_{l}(\mathbf{x},t) = - G\varphi_{l}(\mathbf{x},t) \sum_{l' \neq l} \sum_{i}\omega _{i}\varphi _{l'}(\mathbf{x}+\mathbf{c}_{i},t) \mathbf{c}_{i}  \hspace{.2in}
\end{equation}
where $G$ is a coefficient that regulates the strength of the interactions between the two components. The pseudo-potential $\varphi_{l}$ is set equal to the density,  for the sake of simplicity, i.e. $\varphi_{l}(\mathbf{x},t) = \rho_{l}(\mathbf{x},t)$. When the coupling strength is large enough, the system can show phase segregation with the formation of diffuse interfaces separating bulk regions with majority of one of the two components. Diffuse interfaces display widths of the order of a few grid sizes and a positive surface tension $\gamma$ which increases at increasing $G$. The term $\xi _{li}$ in~\cref{eq:lbe} is a noise term that is assumed to be a zero-mean Gaussian random variable, uncorrelated in time and with constant variance (which can however be space-dependent). While noise does not introduce stochastic forces on the density modes, it does so in momentum modes with the following correlations
\begin{equation}\label{eq:FDT1}
\left \langle \xi _{l n} (\mathbf{x},t)\xi _{l n'} (\mathbf{x}',t') \right \rangle = - \left \langle \xi _{l n} (\mathbf{x},t)\xi _{l' n'} (\mathbf{x}',t') \right \rangle = (2\kappa - \kappa^2) k_{B}T\frac{\rho_{l}\rho_{l'}}{\rho_{l}+\rho_{l'}}\delta_{n n'} \delta(\mathbf{x}-\mathbf{x}') \delta(t-t') \hspace{.2in} n,n' = 1,2,3 \hspace{.2in} l=A,B \hspace{.2in} l\neq l'
\end{equation}
where $n$, $n'$ refer to the modes and $\kappa$ represents the relaxation frequency of the momentum modes in the MRT scheme. The noise correlations on higher modes satisfy
\begin{equation}\label{eq:FDT2}
\left \langle \xi_{l n}(\mathbf{x},t)\xi _{l n'}(\mathbf{x}',t') \right \rangle = (2\kappa - \kappa^2) N_{n} \frac{k_{B}T}{c_s^{2}}\rho_{l}\delta_{n n'} \delta(\mathbf{x}-\mathbf{x}') \delta(t-t')  \hspace{.2in} n,n' = 4,...,18
\end{equation}
where $N_n$ are normalization constants fixed by $N_n \delta_{n n'} =  \sum_{i} w_i \mathbf{e}_{n i}\mathbf{e}_{n' i}$. Notice that all other noise correlation vanish. At hydrodynamical scales, the fluctuating LB allows to obtain -- via the Chapman Enskog analysis~\cite{Dunweg07,SchillerThesis} -- the stochastic hydrodynamic equations for a binary fluid (repeated indexes are meant summed upon)~\cite{Zarate2006}
\be\label{eq:hydro1}
\partial_t \rho_{\st{tot}} + \partial_{\alpha} ( \rho_{\st{tot}} v_{\alpha})=0,\hspace{.2in} \partial_t \rho_A + \partial_{\alpha} (\rho_A v_{\alpha})=\partial_{\alpha} \left[ \mathcal{D} \partial_{\alpha} \mu+\Psi_{\alpha} \right]
\ee
\be\label{eq:hydro2}
\partial_t (\rho_{\st{tot}} v_{\alpha}) + \partial_{\beta} (\rho_{\st{tot}} v_{\alpha} v_{\beta}) = - \partial_{\beta} P_{\alpha \beta} + \partial_{\beta} [\eta (\partial_{\alpha} v_{\beta} + \partial_{\beta} v_{\alpha} ) + \Sigma_{\alpha \beta}].
\ee
The equilibrium properties are fully encoded in the chemical potential $\mu$ and the pressure tensor $P_{\alpha \beta}$ which depend on the interaction model chosen at the level of LB~\cite{SC93,SC94} and whose expressions may be found elsewhere~\cite{SegaSbragaglia13}. The terms $\Psi_{\alpha}$ and $\Sigma_{\alpha \beta}$ are stochastic fluxes and tensors, respectively. Specifically, the stochastic vector field $\Psi_{\alpha}$ is the term due to the thermal noise that must be added  to the diffusion flux $\mathcal{D} \partial_{\alpha} \mu$~\cite{Zarate2006}, with $\mathcal{D}$ the diffusion constant; the stochastic tensor $\Sigma_{\alpha \beta}$ is added to the viscous stress tensor $\eta (\partial_{\alpha} v_{\beta} + \partial_{\beta} v_{\alpha})$~\cite{Landau}, with $\eta$ the dynamic viscosity for the bulk. Requiring that the fluctuation-dissipation relation holds for our hydrodynamical problem, and using \eqref{eq:FDT1} and \eqref{eq:FDT2}, one can derive  a unique choice for the intensity of the stochastic contributions~\cite{Zarate2006,Belardinelli15}:
\be
\Sigma_{\alpha \beta}=\sqrt{\eta k_B T}(W_{\alpha \beta}+W_{\beta \alpha}^T) \hspace{.2in} \Psi_{\alpha}=\sqrt{2 \mathcal{D}  k_B T} \tilde{W}_{\alpha}
\ee
where $k_B$ is the Boltzmann constant, and $W_{\alpha \beta}$ and $\tilde{W}_{\alpha}$ are random Gaussian tensors and a random Gaussian vector field respectively, with independent components and variance equal to unity. 


\section{Numerical Set-up and results}\label{sec:ligamentsetup}
Numerical simulations are conducted in a 3D fully periodic domain with sizes $L_x \times L_y \times L_z$. A cylindrical ligament with a majority of phase A and radius $R_0$ is set-up with a symmetry axis along the $z$ coordinate (see \cref{fig:ligament}). By keeping fixed the ratio between the system sizes $L_{x,y,z}$ and the ligament initial radius $R_0$, we have performed different numerical simulations at changing the thermal length $\ellT$ and the domain resolution $L_x \times L_y \times L_z$. For each realization of the thermal length and domain resolution, we performed hundreds of simulations to gather sufficient statistics over the break-up time and the droplet volumes after break-up. For computational reasons, larger resolutions are associated with a smaller number of simulations. All these parameters are summarized in \cref{tab:parameter_space}. 
\begin{table}[H]
\centering
\begin{tabular}{ @{} c  c c c c c c c @{} }
\toprule[1pt]
$R_{0}$ (lbu) & $L_x$ (lbu) & $L_y$ (lbu) & $L_z$ (lbu) & $\ell_{\mbox{\tiny T}}^{2}$ (lbu$^2$) & $\#\, \mbox{of simulations}$ \\
\midrule[0.5pt]
\addlinespace[0.5mm]
7.0 & 48   & 48   & 128 & $ 7 \cdot10^{-5} - 3\cdot10^{-3} $ & $1000$ \\
10.0 & 72   & 72   & 180 & $1\cdot10^{-4} - 3\cdot10^{-3}$ & $500$   \\
14.0 & 96   & 96   & 256 & $1\cdot10^{-4}$ & $200$   \\
18.0 & 122 & 122 & 324 & $1\cdot10^{-4}$ & $200$   \\
28.0 & 192 & 192 & 512 & $1\cdot10^{-4}$ & $100$   \\
\bottomrule[1pt]
\end{tabular}
\caption{Summary of the different numerical simulations conducted. All the numerical simulations that we describe in this paper are performed with a coupling coefficient $G = 1.5$ LB units (lbu hereafter) in \cref{sc-force_eq}.  Inside the ligament, the density for the two components are set to $\rho_A= 2.21$ lbu and $\rho_B = 0.09$ lbu, with a corresponding total density $\rho_{\st{tot}} = 2.3$ lbu. The surface tension of the system is $\gamma = 0.1515$ lbu. The surface tension is kept fixed in all the numerical simulations, while the noise intensity is varied in equations \eqref{eq:hydro1}-\eqref{eq:hydro2} to achieve different thermal lengths. The viscosity ratio between the dispersed and continuous phases is set equal to unity. }
\label{tab:parameter_space}
\end{table}
We remark that applications of LB in the problem of ligament contraction have already been proposed in the literature. In particular, in \cite{srivastava2013} a comparison among  axisymmetric LB and the predictions of deterministic sharp interface hydrodynamics has been provided. The use of an axisymmetric LB obviously reduces the computational effort; however, in presence of thermal noise, it does not provide a realistic description of the interfacial fluctuations typical of 3D interfaces~\cite{Grant,Safran}. For this reason we have used a fully 3D fluctuating LB without any axial symmetry. In all the simulations that we conducted, the length of the domain size is chosen to be $L_x \approx 18\, R_{0}$, which is well suited to accommodate roughly 2 wavelengths $\lambda_{\mbox{\tiny fast}}$ of the fastest-growing mode of the Plateau-Rayleigh instability \cite{eggers1994drop} (see also \cref{fig:ligament} for quantitative details). The numerical simulations are then conducted for a set of parameters for which the Ohnesorge number $\mbox{Oh}$  is small, $(\mbox{Oh} < 1)$. The Ohnesorge number quantifies the importance of the viscous forces with respect to the inertial and surface tension forces, and is defined as $\mbox{Oh}=\eta/\sqrt{\rho^{\mbox{\tiny max}}_A \gamma R_0}$, where $\rho^{\mbox{\tiny max}}_A$ indicates the maximum density of the ligament and $\eta$ is the dynamic viscosity. The fastest-growing mode has a wave-number $k_{\mbox{\tiny fast}} R_0 \approx 0.697$ , where $\lambda_{\mbox{\tiny fast}} = k_{\mbox{\tiny fast}}^{-1}$,  independently on $\mbox{Oh}$ \cite{eggers1994drop}.\\

\subsection{Effects of thermal fluctuations}\label{sec:PRI} 
\begin{figure}[H]
\centering
\includegraphics[width=1\textwidth]{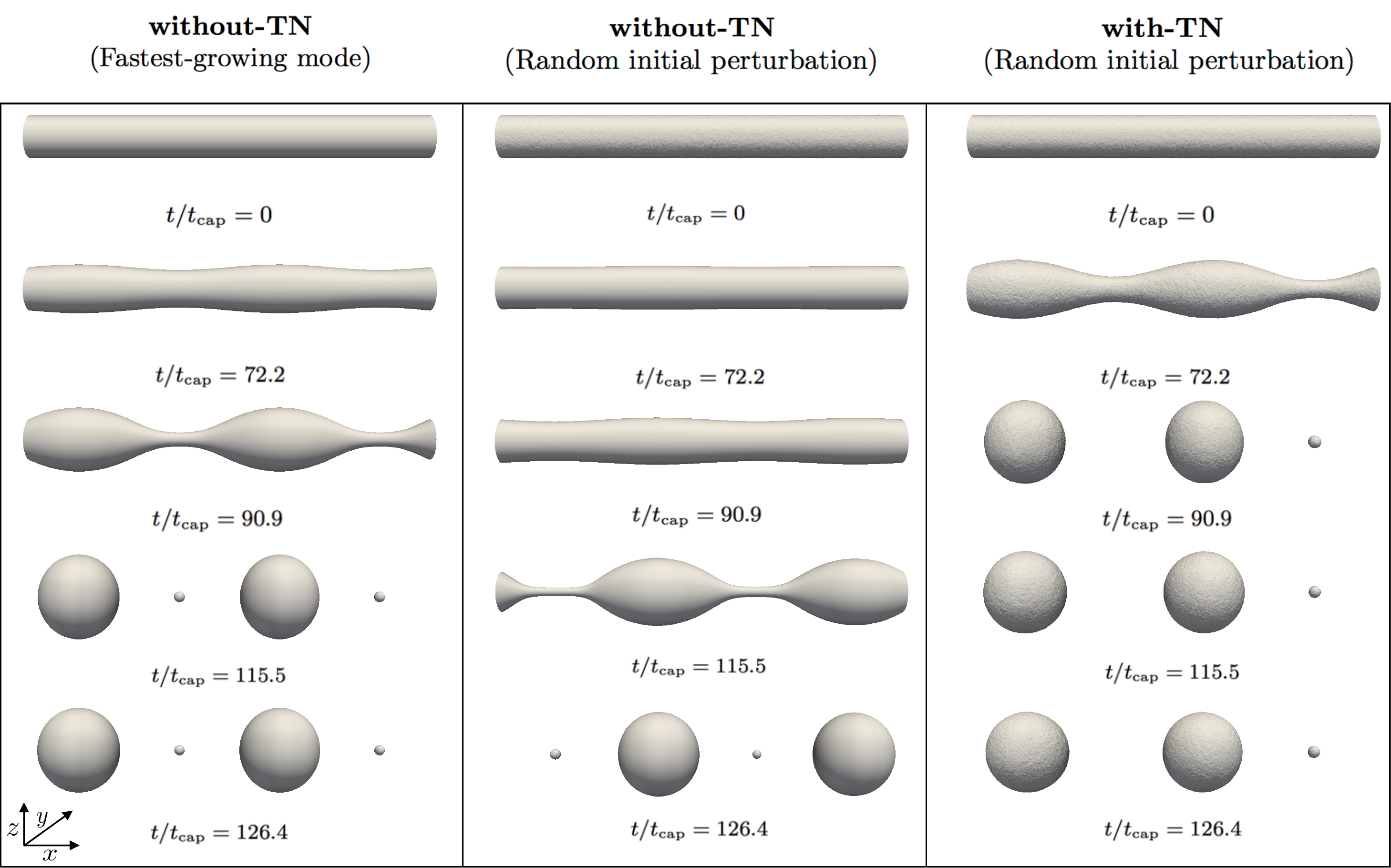}
\caption{Ligament break-up with LB simulations. Left column panels: LB deterministic evolution of a liquid ligament destabilized by the Plateau\texttt{-}Rayleigh instability. The system is initialized with a small perturbation on the fastest growing mode with wavelength $\lambda_{\mbox{\tiny fast}}$. The axial system size of the numerical simulations is chosen to accommodate 2 $\lambda_{\mbox{\tiny fast}}$. After the break-up, we observe that the volumes of the two ``mother'' droplets are equal to each other, the same happens for the ``satellite'' droplets. Middle column panels (``without-TN" protocol): random initial condition followed by LB deterministic evolution (see text for details on preparation). Right column panels (``with-TN" protocol): random initial condition followed by fluctuating hydrodynamics evolution with thermal noise (details are reported in the text).}
\label{fig:ligament}
\end{figure}

We are interested to understand what is the effect of the thermal noise on the ligament break-up in combination with the Plateau\texttt{-}Rayleigh instability. To this aim we have designed three different simulation cases, as shown in \cref{fig:ligament}. In the first case (left panels), we evolve the ligament without thermal noise, i.e. we use Eqs. \eqref{eq:hydro1}-\eqref{eq:hydro2} with $k_BT=0$. We set an initial small perturbation with wavelength $\lambda_{\mbox{\tiny fast}}$ and very small amplitude. Due to the Plateau\texttt{-}Rayleigh instability, the unstable mode along the interface grows and eventually determine the break-up of the ligament into droplets. In the second case (middle panels), we evolve the same equations with a different initial condition: beyond the perturbation on the fastest growing mode, we also add a random Gaussian perturbation on the less unstable Fourier modes; the evolution is kept deterministic, i.e. again we use Eqs. \eqref{eq:hydro1}-\eqref{eq:hydro2} with $k_BT=0$. In the third case (right panels), we show results coming from numerical simulations with the same initial condition used for the middle panels followed by the fluctuating hydrodynamics evolution, i.e. Eqs. \eqref{eq:hydro1}-\eqref{eq:hydro2} with $k_BT>0$. Typically, after a few capillary times $\tcap = \sqrt{\rho^{\mbox{\tiny max}}_A R_0^{3}/ \gamma}$ the ligament breaks into two ``mother'' droplets and two ``satellite'' droplets. However, in the case of thermal fluctuations (right panel in \cref{fig:ligament}), the enhanced volume polydispersity may cause one of the satellite droplets to be so small that it cannot be resolved with the resolution used. The presence of two mother droplets is clearly due to the fact that we choose an axial length that corresponds to twice the wavelength of the fastest growing mode. The presence of small satellite droplets is generated by the combined effect of viscosity and surface tension at the late stage of pinch-off, as already described in the literature \cite{rutland1970, Eggers97, ashgriz1995}. This qualitative feature is robust and independent of the thermal noise and of the initialization protocol used. For the pure deterministic case (left panels), the fragmentation process evolves in a symmetric way, and it leads to two identical mother droplets and two identical satellite droplets. However, when we add noise either in the initial configuration or during the entire evolution, things become more complicated: the break-up time is a random variable and a volume polydispersity in both mother and satellite droplets is observed. In particular, the presence of thermal noise in the evolution (right panels) manifestly accelerates the break-up, which is consistent with previous studies \cite{Moseler00, Eggers02}. This is because the effects of thermal fluctuations dominate at the late stage of pinch-off regime. It is apparent from \cref{fig:ligament} that during the ligament fragmentation process both the random initial conditions and thermal fluctuations have a role; more importantly, numerical simulations offer the possibility to quantitatively disentangle the two contributions by using two complementary simulations protocols: we can change the random initial condition and integrate a deterministic dynamics without thermal noise ($k_BT=0$) in Eqs. \eqref{eq:hydro1}-\eqref{eq:hydro2} (``without-TN'' protocol, as in middle panels of \cref{fig:ligament}) or we can include the effects of thermal noise in the dynamics by using $k_BT \neq 0$ in Eqs. \eqref{eq:hydro1}-\eqref{eq:hydro2} (``with-TN'' protocol, as in right panels of \cref{fig:ligament}). Based on these two simulation protocols, in the following we aim at characterizing the statistics of the droplet break-up times and the droplet volume distribution. The quantitative characterization of the droplet volume statistics naturally poses the question of the specificity of the results, due to the fact that LB is a diffuse interface hydrodynamics solver. To address this point we will perform a quantitative comparison between the droplet volume statistics from the simulations and the predictions of sharp interface hydrodynamics.


\subsection{Statistics of break-up times}\label{sec:breakup}

\begin{figure}[H]
\begin{minipage}{1\textwidth}
\centering
\subfigure{\includegraphics[width = 0.32\linewidth]{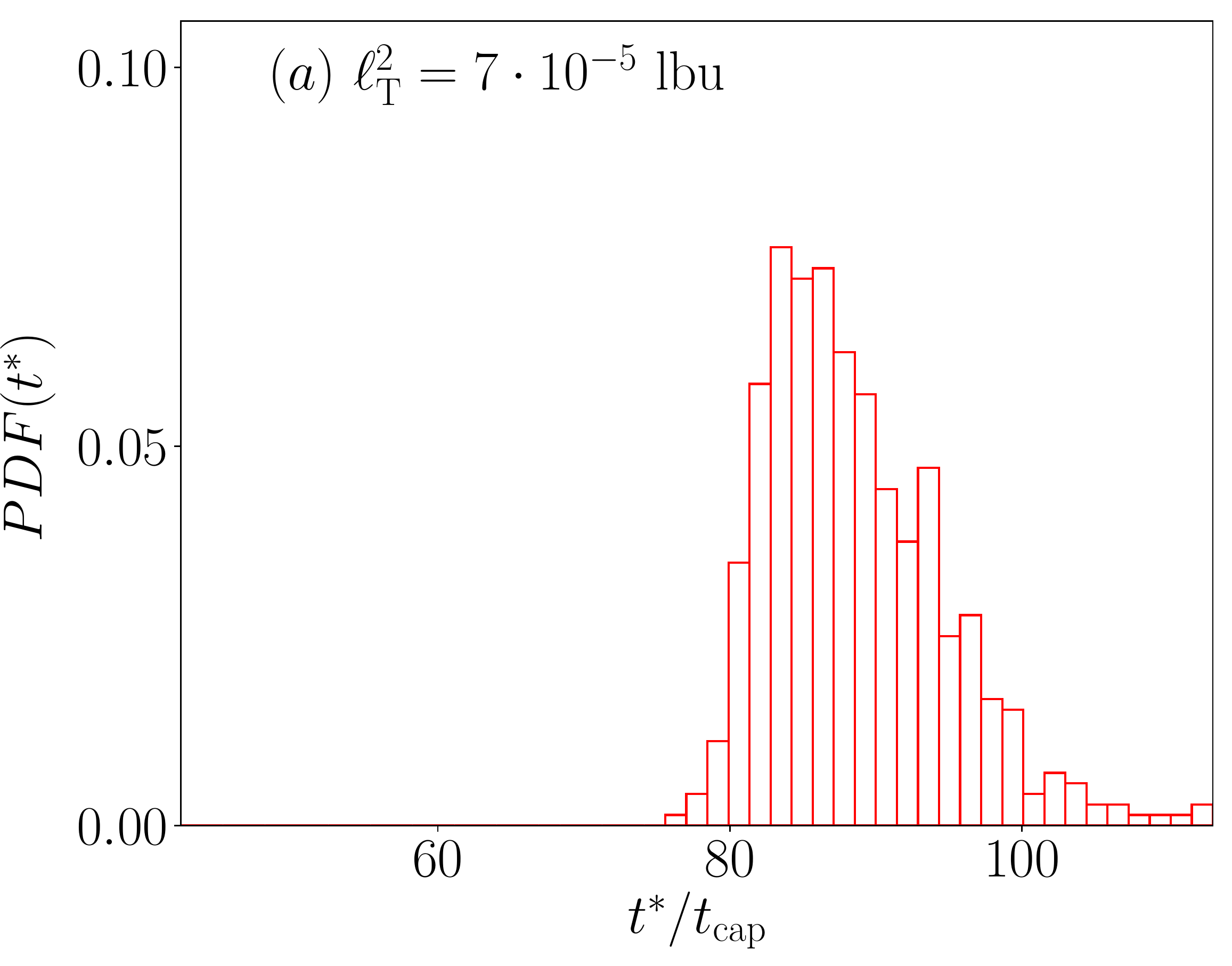}\label{fig:tbLBMPDF1}}
\subfigure{\includegraphics[width = 0.32\linewidth]{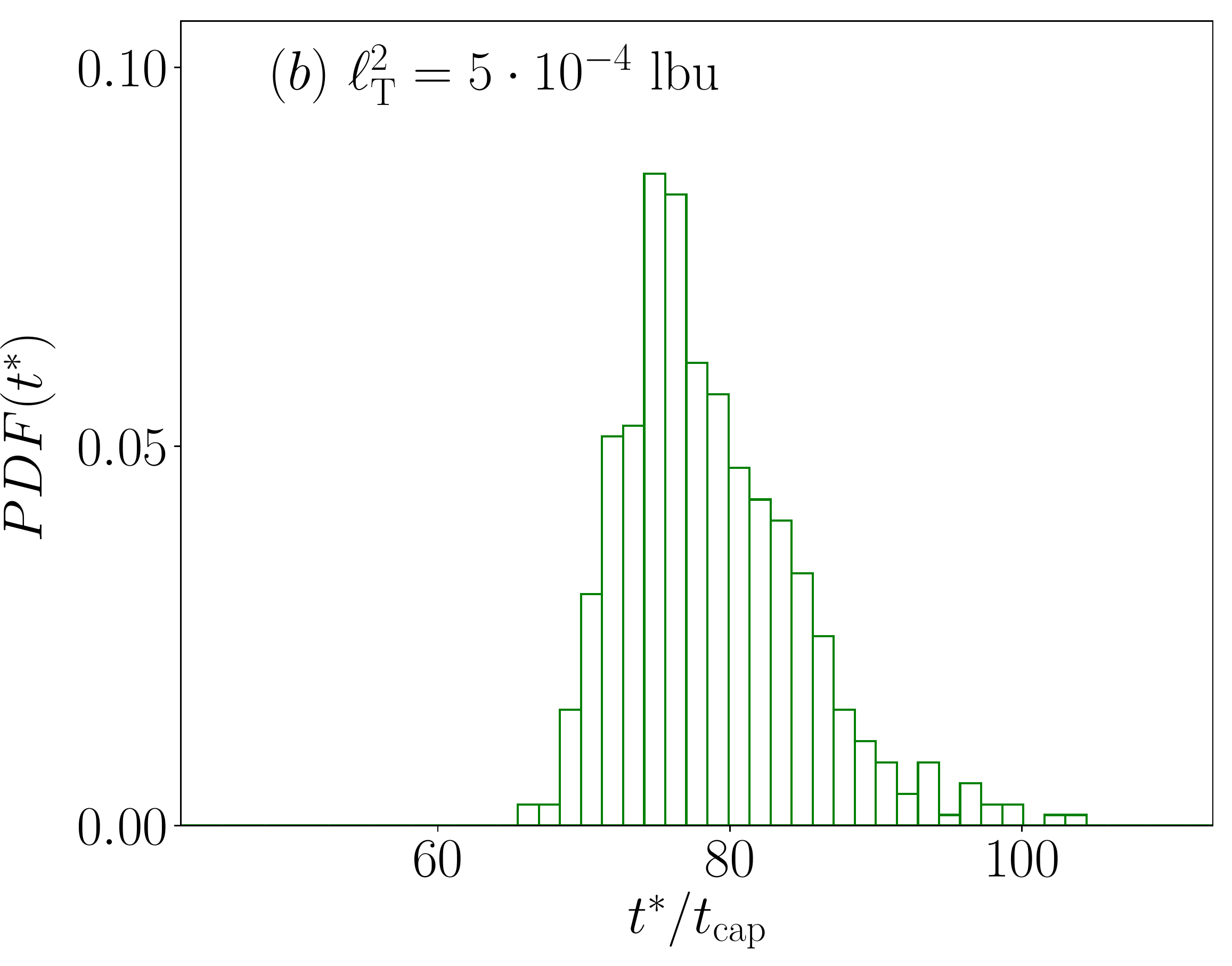}\label{fig:tbLBMPDF2}}
\subfigure{\includegraphics[width = 0.32\linewidth]{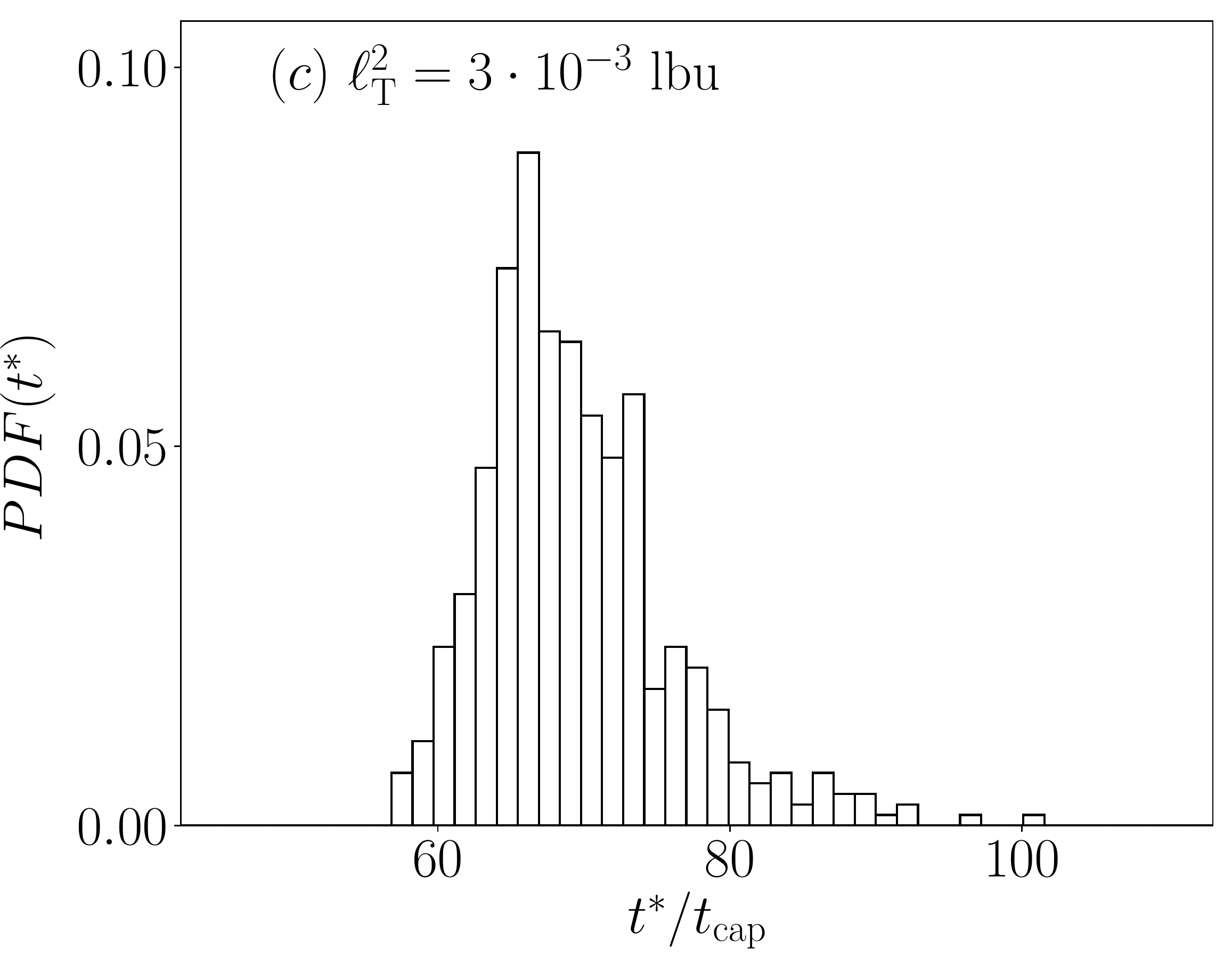}\label{fig:tbLBMPDF3}}\\
\end{minipage}\\
\begin{minipage}{1\textwidth}
\centering
\subfigure{\includegraphics[width = 0.32\linewidth]{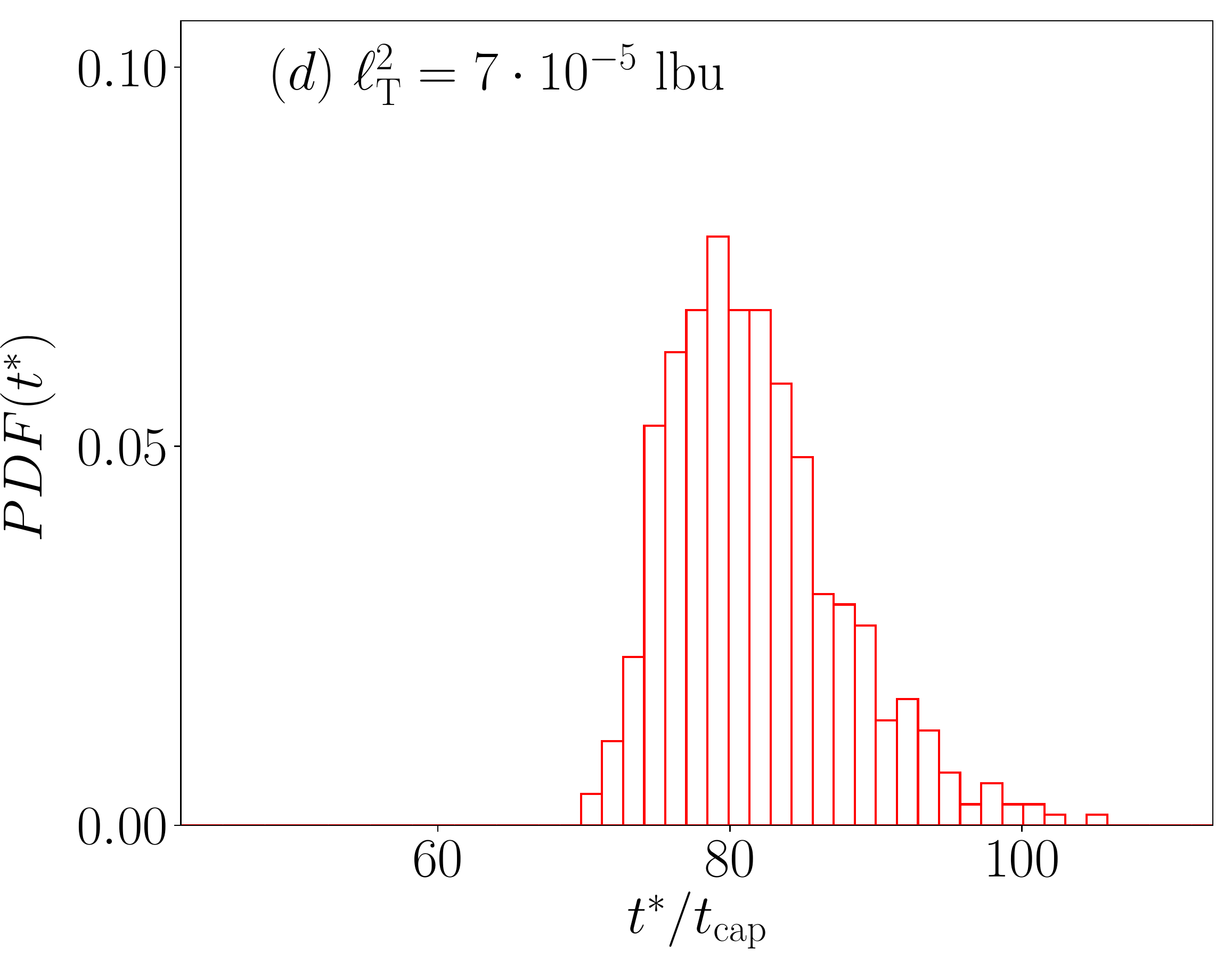}\label{fig:tbFLBMPDF1}}
\subfigure{\includegraphics[width = 0.32\linewidth]{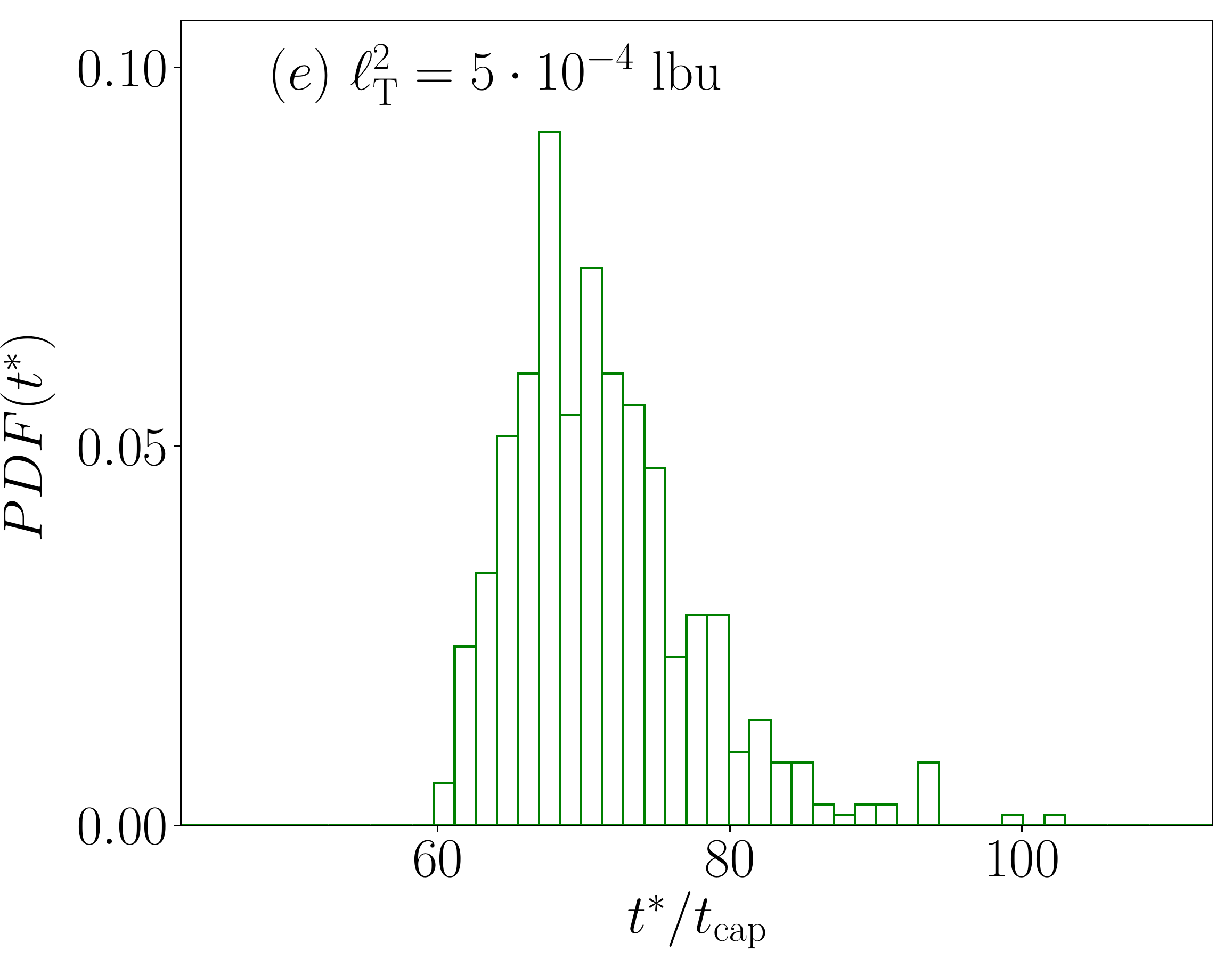}\label{fig:tbFLBMPDF2}}
\subfigure{\includegraphics[width = 0.32\linewidth]{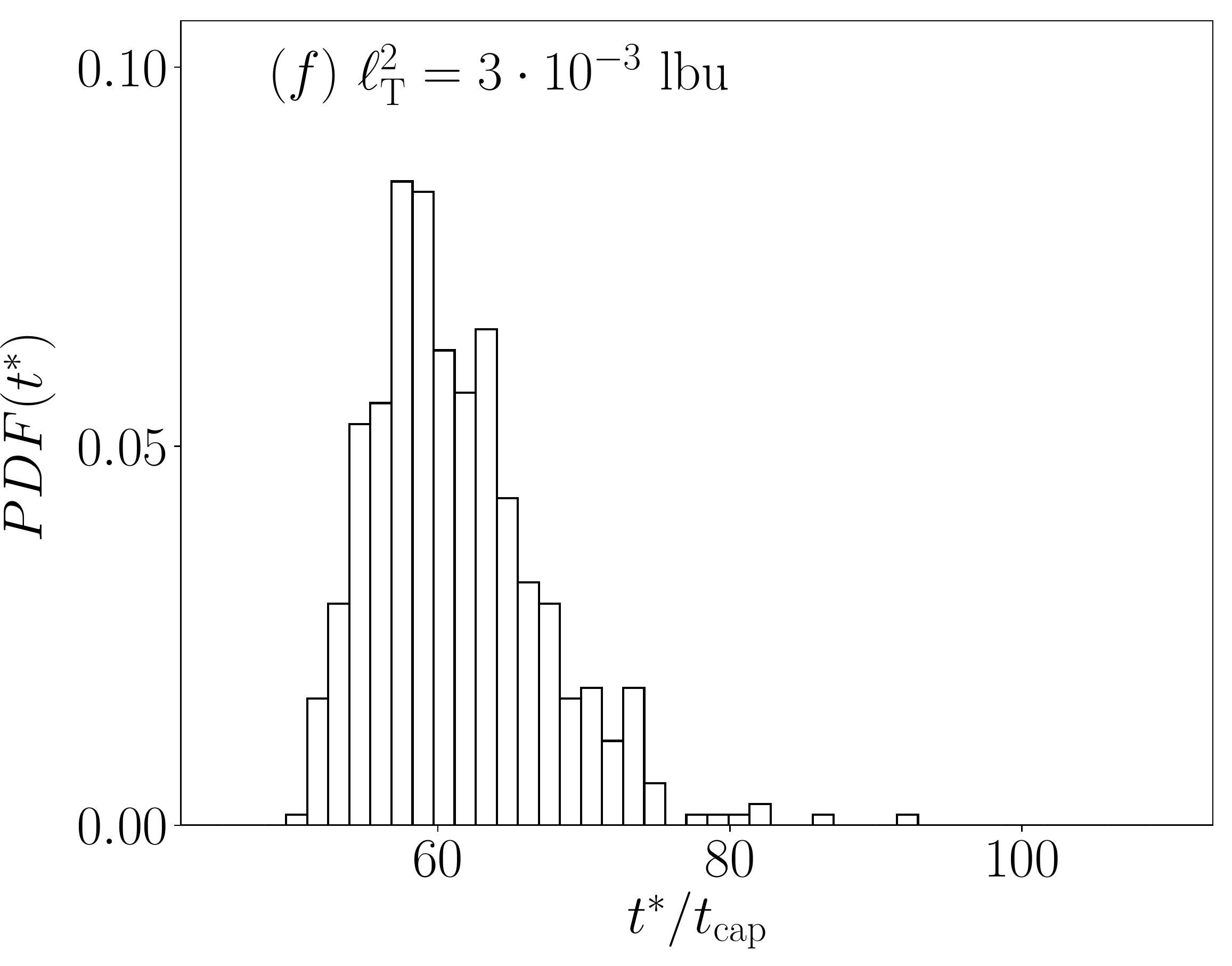}\label{fig:tbFLBMPDF3}}\\
\end{minipage}
\caption{PDFs for the break-up time $t^*$ as a function of the thermal length squared $\ellT^2$ for different simulations protocols (see \cref{fig:ligament}): ``without-TN'' protocol (Panels (a)-(c)) and ``with-TN'' protocol (Panels (d)-(f)). The break-up time is made dimensionless with the capillary time $\tcap$.} 
\label{fig:tbreakPdf}
\end{figure}

Based on previous experimental and numerical works~\cite{Hennequin06,Petit12, Eggers02}, we know that the thermal noise results in accelerating the late stage of the pinch-off process. Beyond this accelerated dynamics, here we focus on characterizing the statistics of the break-up times. We have conducted numerical simulations for a fixed ligament radius $R_0=7$ lbu and different thermal lengths, $\ellT$ ranging from $7 \cdot 10^{-5}$ to $3 \cdot 10^{-3} $ lbu, in the two simulation protocols ``with-TN" and ``without-TN''. To measure the break-up time, we follow the ligament fragmentation evolution and record the time when a discontinuity in the density profile is observed along the ligament axis. In~\cref{fig:tbreakPdf}, we study the probability density function (PDF) of the break-up time $t^{*}$. Overall, we observe that the shapes of the PDFs are similar to the break-up time distribution reported in a previous fluctuating thin films study~\cite{grun2006thin}. Comparing the ``without-TN'' protocol (first row) and the ``with-TN'' protocol (second row) for each thermal length, we see that the thermal fluctuations increase the probability for the ligament to break-up sooner, which makes the peaks of the distribution of $t^{*}$ moving closer to the origin. Similarly, for both simulation protocols, we find a systematic speed-up of the break-up time by increasing the amplitude of the thermal fluctuations, as shown by comparing PDFs on the same row at increasing thermal length (from left to right). 

\begin{figure}[H]
\begin{minipage}{1.0\textwidth}
\centering
\subfigure{\includegraphics[height = 0.38\linewidth] {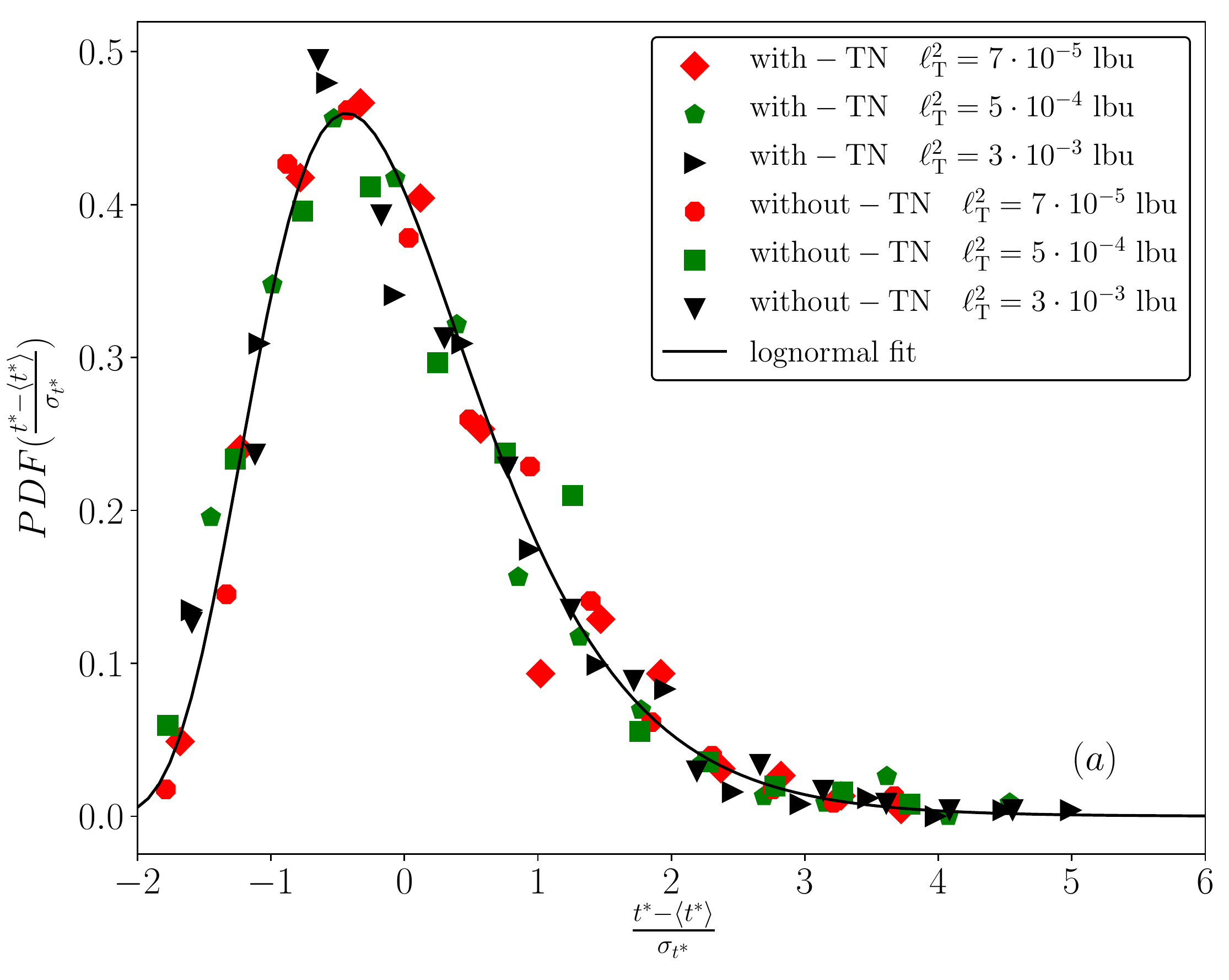}\label{fig:normalBreakPdf}}
\subfigure{\includegraphics[height = 0.38\linewidth]{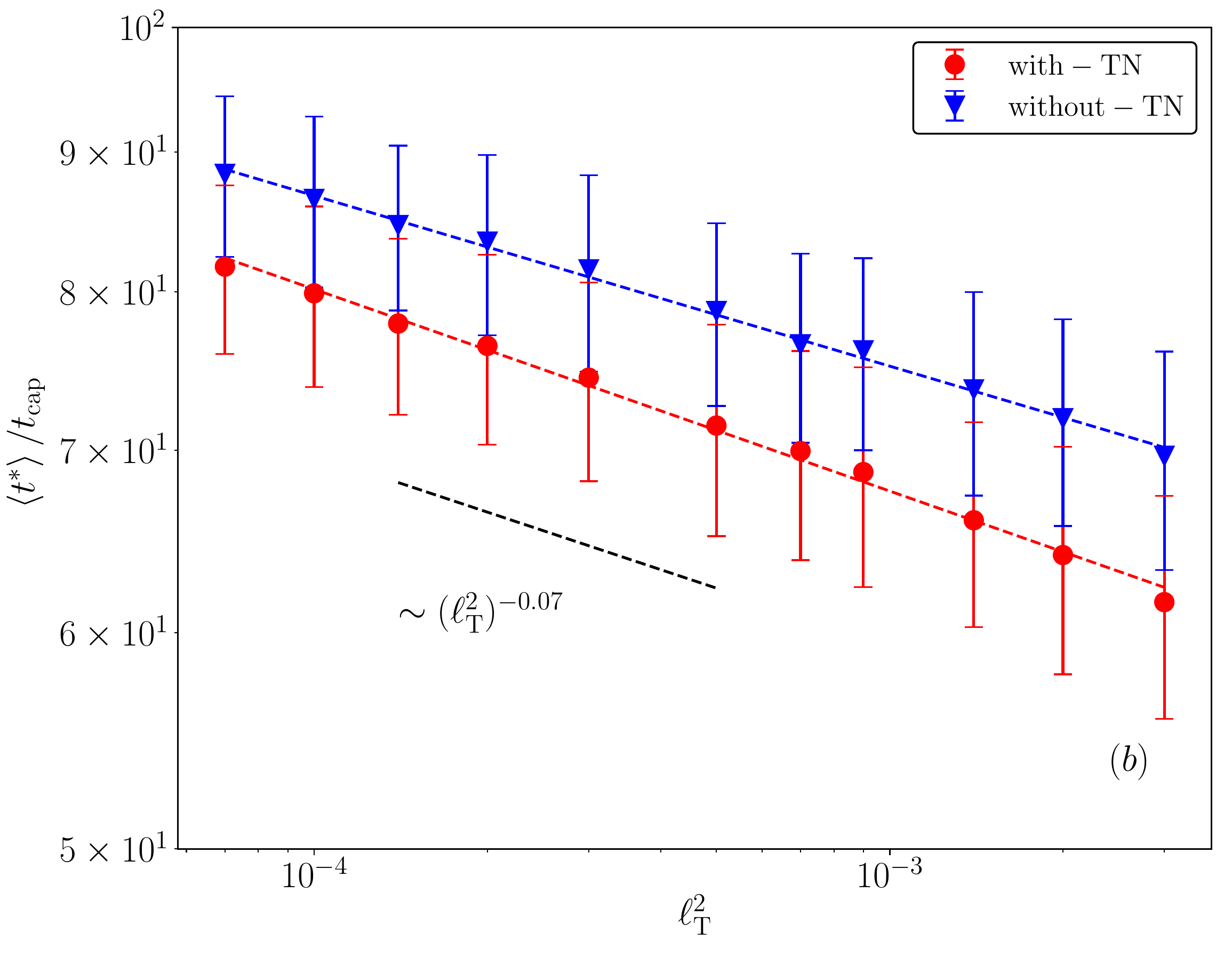}\label{fig:tbreakLaw}}\\
\end{minipage}
\caption{Panel (a): PDF of normalized break-up time at changing thermal length $\ellT$ for fixed ligament radius $R_0$ for both simulation protocols fitted with lognormal distribution. Panel (b): dimensionless mean break-up time $\left \langle t^{*} \right \rangle/\tcap$ as a function of thermal length square  for the two different simulation protocols used (see \cref{fig:ligament}). Red and blue dotted lines are power fits for the dimensionless mean break-up time. Error bars are estimated from the standard deviation (see \cref{fig:tbreakPdf}).}
\label{fig:tbreakSum}
\end{figure}

In \cref{fig:normalBreakPdf}, we show the PDFs of $t^{*}$ for both protocols normalized by their mean $\left \langle t^{*} \right \rangle$ and the standard deviation $\sigma_{t^{*}}$. The break-up time data for both protocols follow similar trends which are well reproduced by the log-normal fit
$$
f_{\mbox{\tiny log}}(x)=\frac{1}{(x-x_0)\sigma_{\mbox{\tiny log}} \sqrt{2 \pi}} e^{-(\log(x-x_0)-\mu)^2/2 \sigma_{\mbox{\tiny log}}^{2}}
$$
with $\sigma_{\mbox{\tiny log}}=0.32$, $\mu=2.82$ and $x_{0}=-2.97$. Also, \cref{fig:tbreakLaw} presents results for the average break-up time $\langle t^* \rangle$ as a function of the thermal length squared $\ell_{\mbox{\tiny T}}^{2}$ for both protocols. We observe a power-law like behaviour
$$
\langle t^*/\tcap \rangle \sim (\ell_{\mbox{\tiny T}}^2)^{-0.07}
$$ 
in both cases ``with-TN'' and ``without-TN'', with the latter case always systematically below the former. 
\subsection{Droplet volumes}\label{sec:PDFV}
\begin{figure}[H]
\begin{minipage}{1\textwidth}
\centering
\subfigure{\includegraphics[height = 0.25\linewidth]{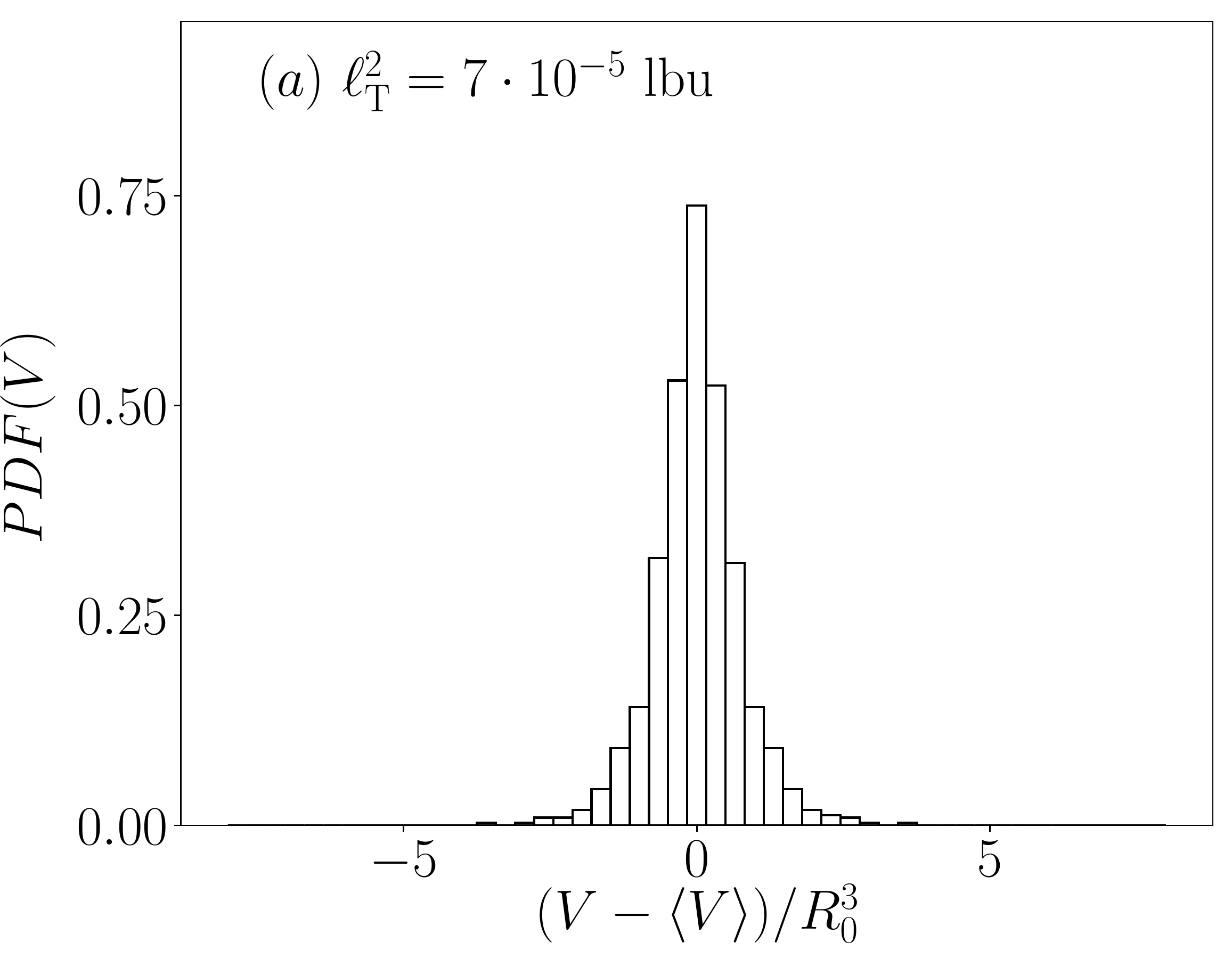}}
\subfigure{\includegraphics[height = 0.25\linewidth]{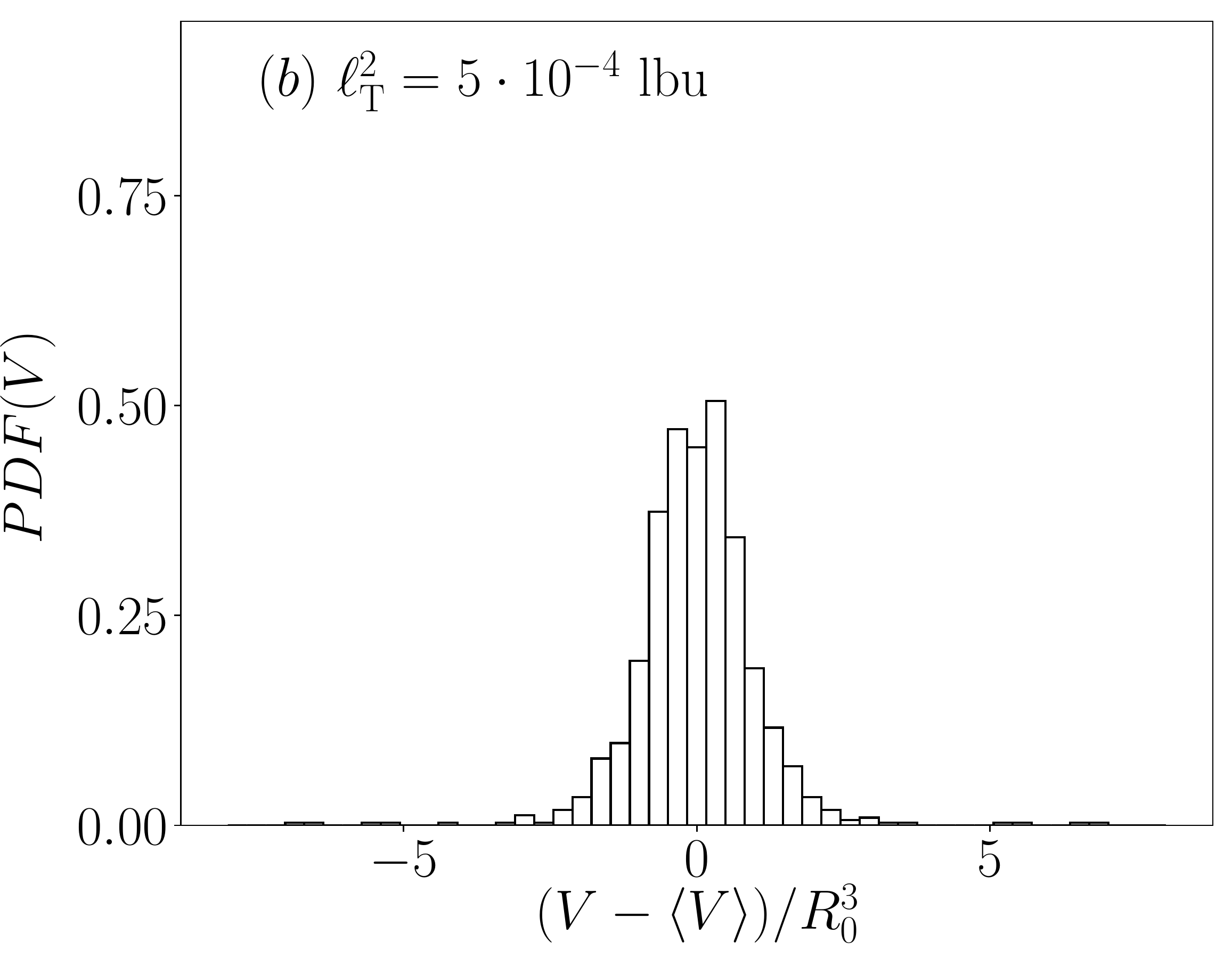}}
\subfigure{\includegraphics[height = 0.25\linewidth]{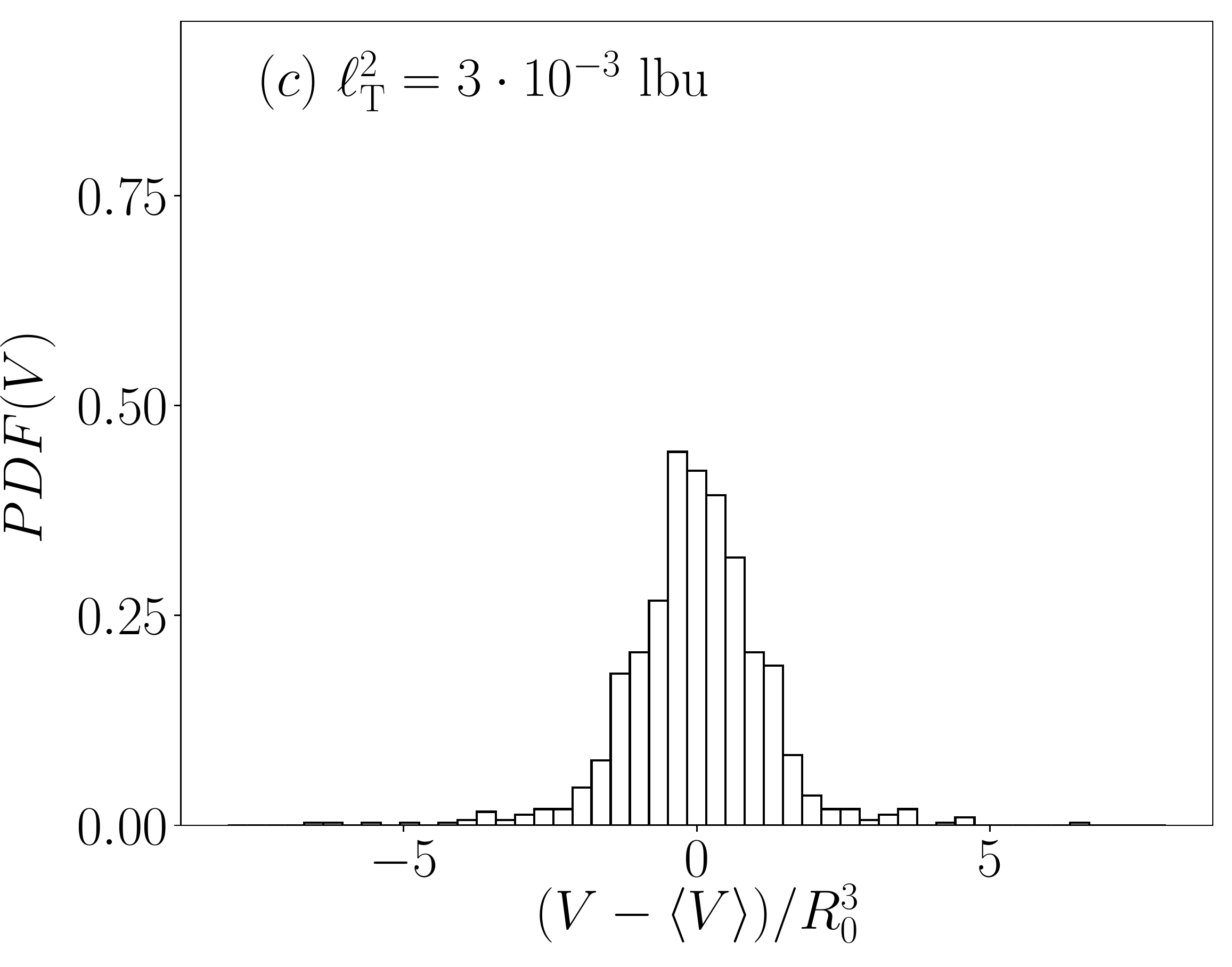}}
\end{minipage}\\

\begin{minipage}{1\textwidth}
\centering
\subfigure{\includegraphics[height = 0.25\linewidth]{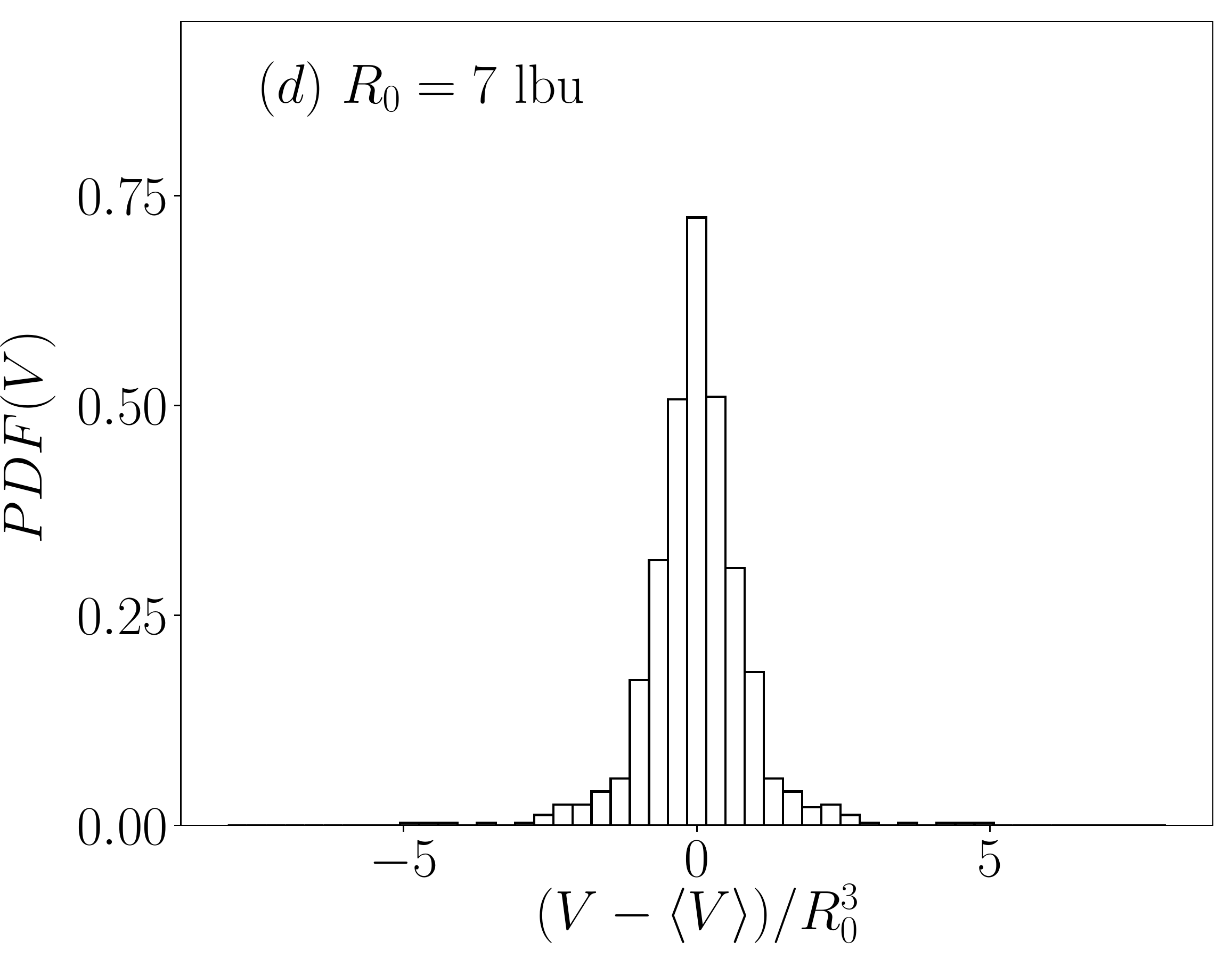}\label{fig:pdf128}}
\subfigure{\includegraphics[height = 0.25\linewidth]{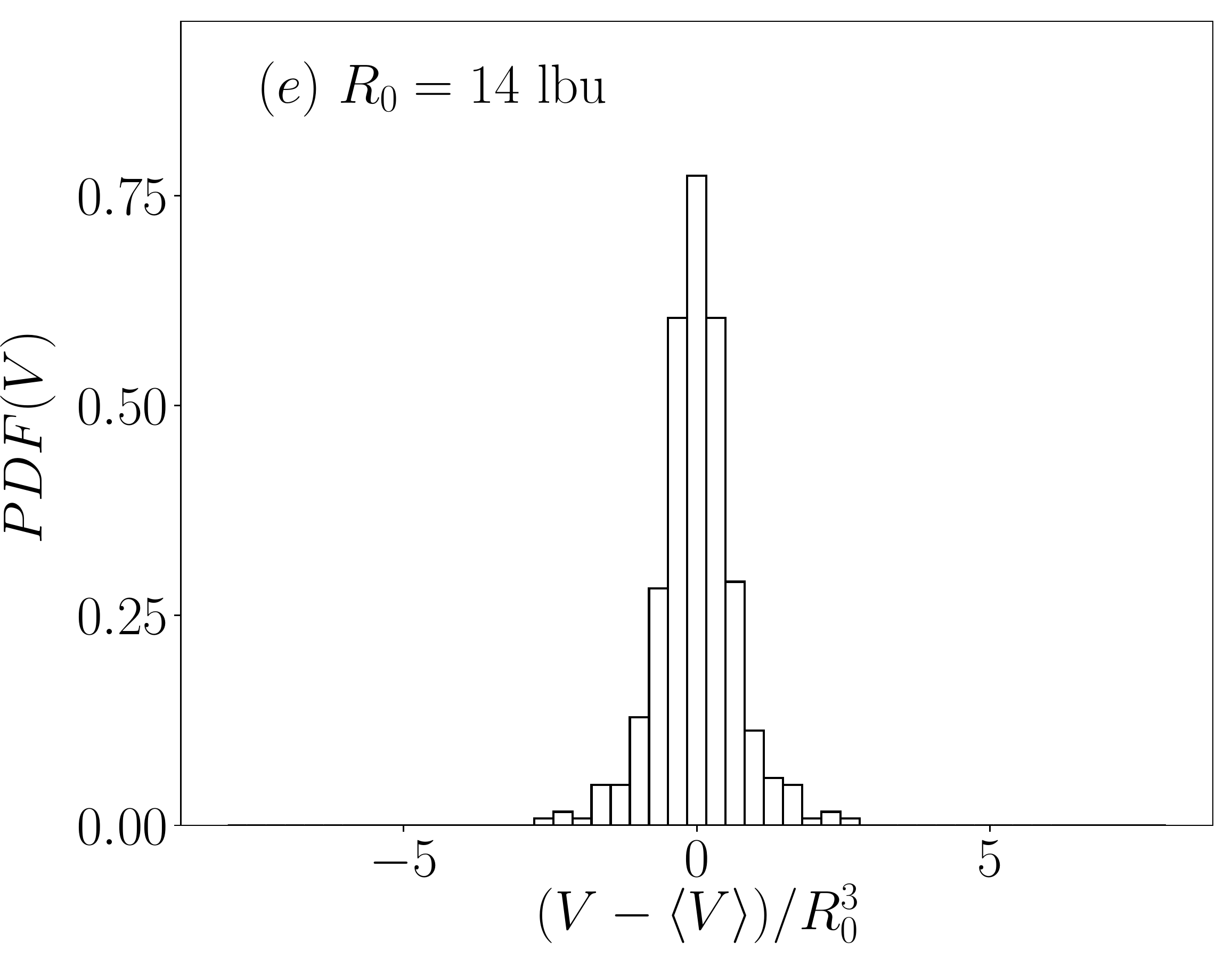}\label{fig:pdf256}}
\subfigure{\includegraphics[height = 0.25\linewidth]{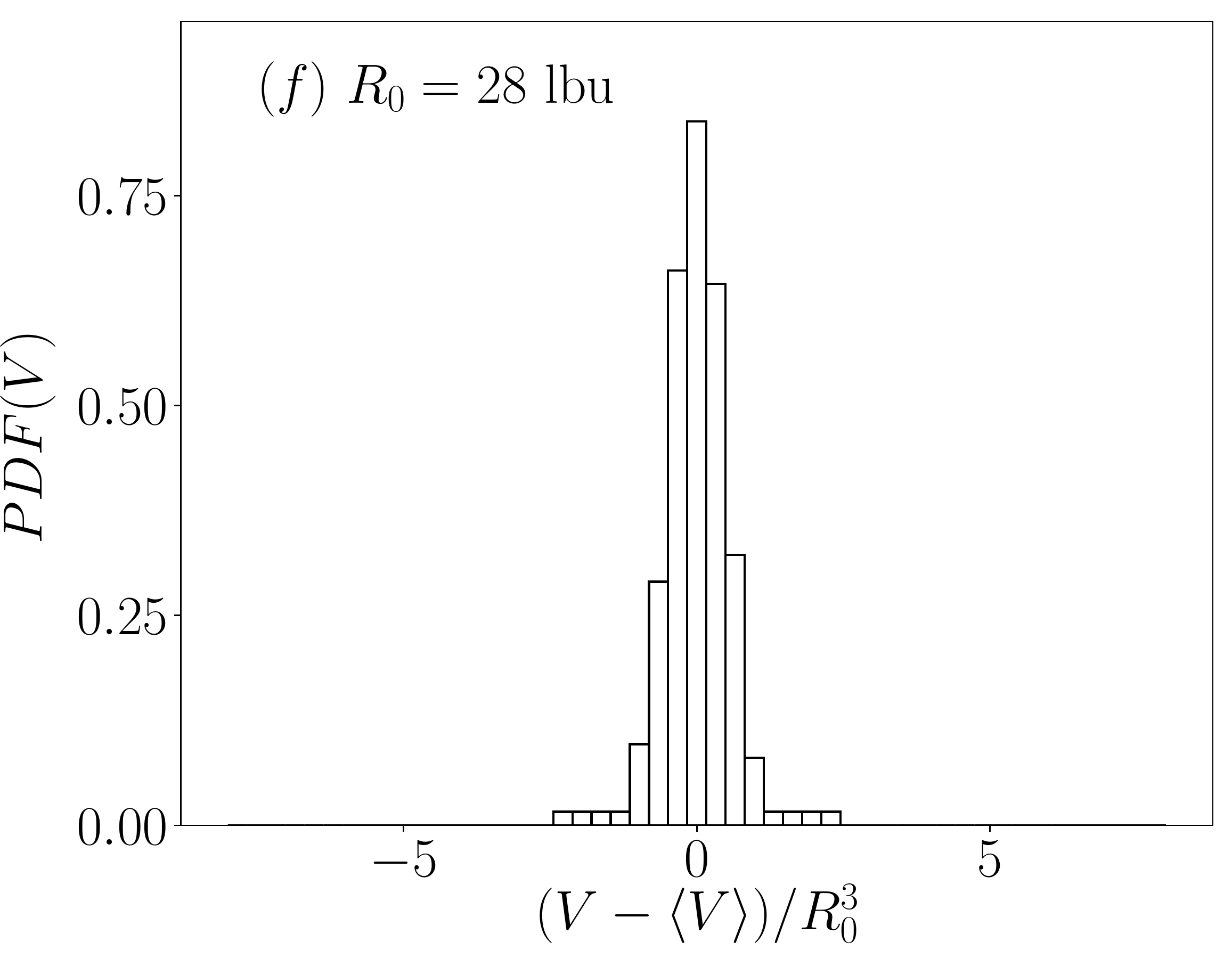}}
\end{minipage}
\caption{PDFs for the droplets volumes after break-up based on the ``with-TN'' protocol (see \cref{fig:ligament}) at changing the thermal length $\ellT$ for fixed ligament radius $R_0 = 7.0$ (top row, panels (a)-(c)) or at changing the ligament radius $R_0$ for fixed thermal length square $\ellT^{2} =  1\cdot10^{-4}$ (bottom row, panels (d)-(f)). To make figures comparable at changing $R_0$, we have subtracted the average volume and rescaled by $R_0^3$.}
\label{fig:volumeCompare}
\end{figure}

\begin{figure}[H]
\begin{minipage}{1.0\textwidth}
\centering
\subfigure{\includegraphics[height = 0.38\linewidth] {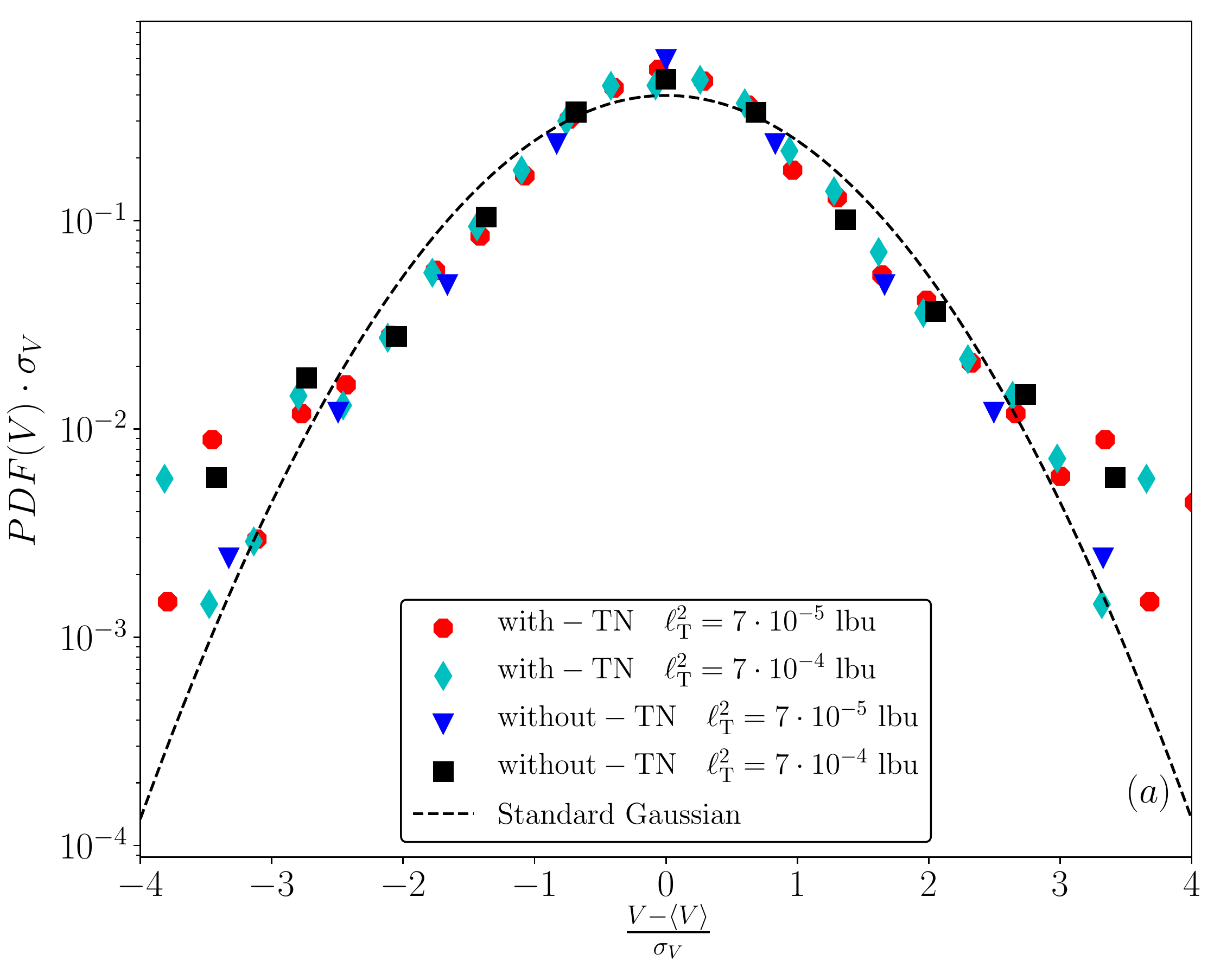}\label{fig:normalPdf}}
\subfigure{\includegraphics[height = 0.38\linewidth]{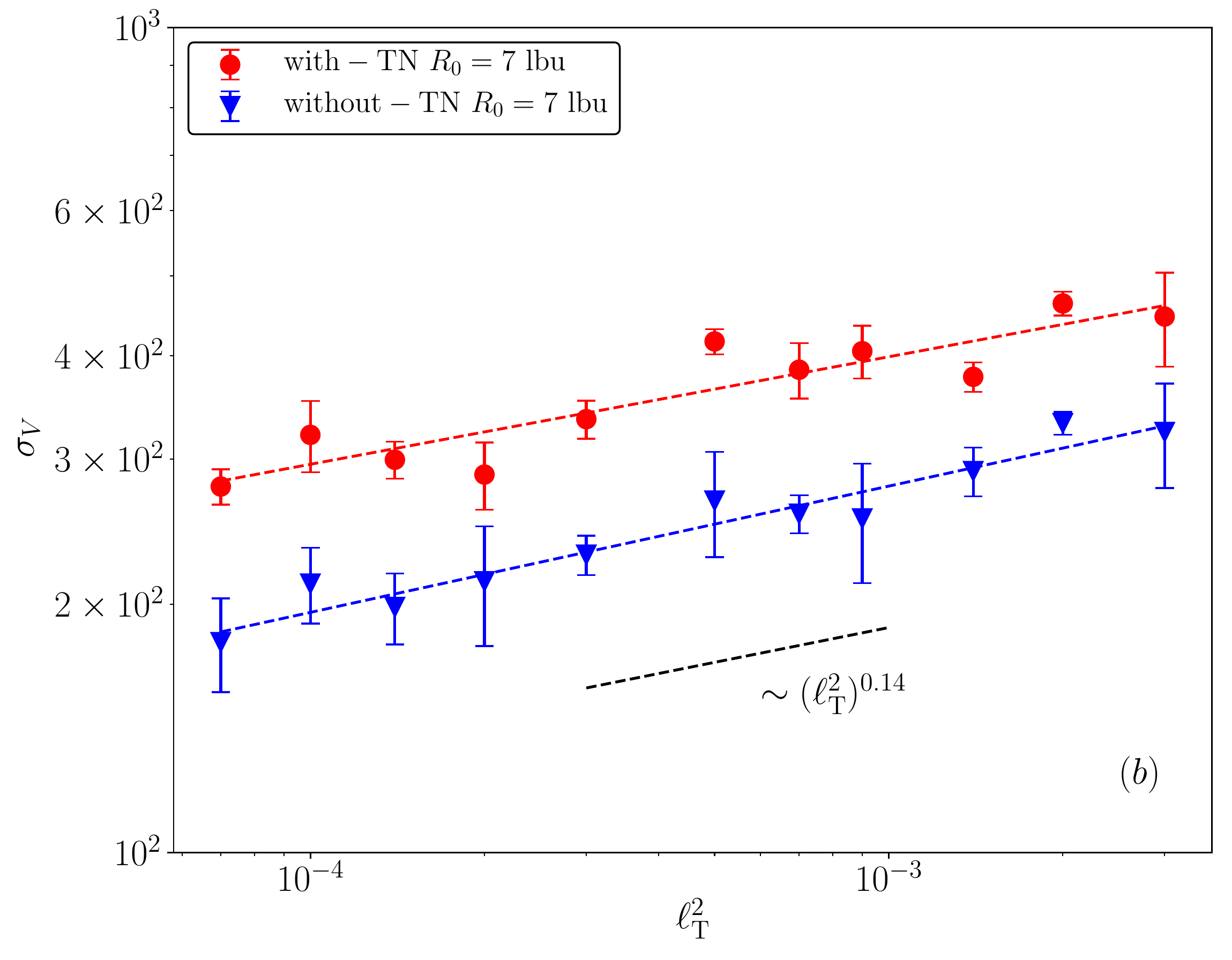}\label{fig:STDlaw}}\\
\end{minipage}
\caption{Panel (a): standardized PDFs at changing the thermal length $\ellT$ at fixed ligament radius $R_0$ for both simulation protocols (``with-TN'' and ``without-TN''). Panel (b): standard deviation, $\sigma_V$, as a function of the thermal length squared $\ellT^2$ at fixed ligament radius $R_0 = 7$ for both simulation protocols. Error bars are estimated from the standard deviation of different groups of the configurations.}
\label{fig:PDFVolumes}
\end{figure}

\begin{figure}[H]
\begin{minipage}{1.0\textwidth}
\subfigure{\includegraphics[height = 0.38\linewidth]{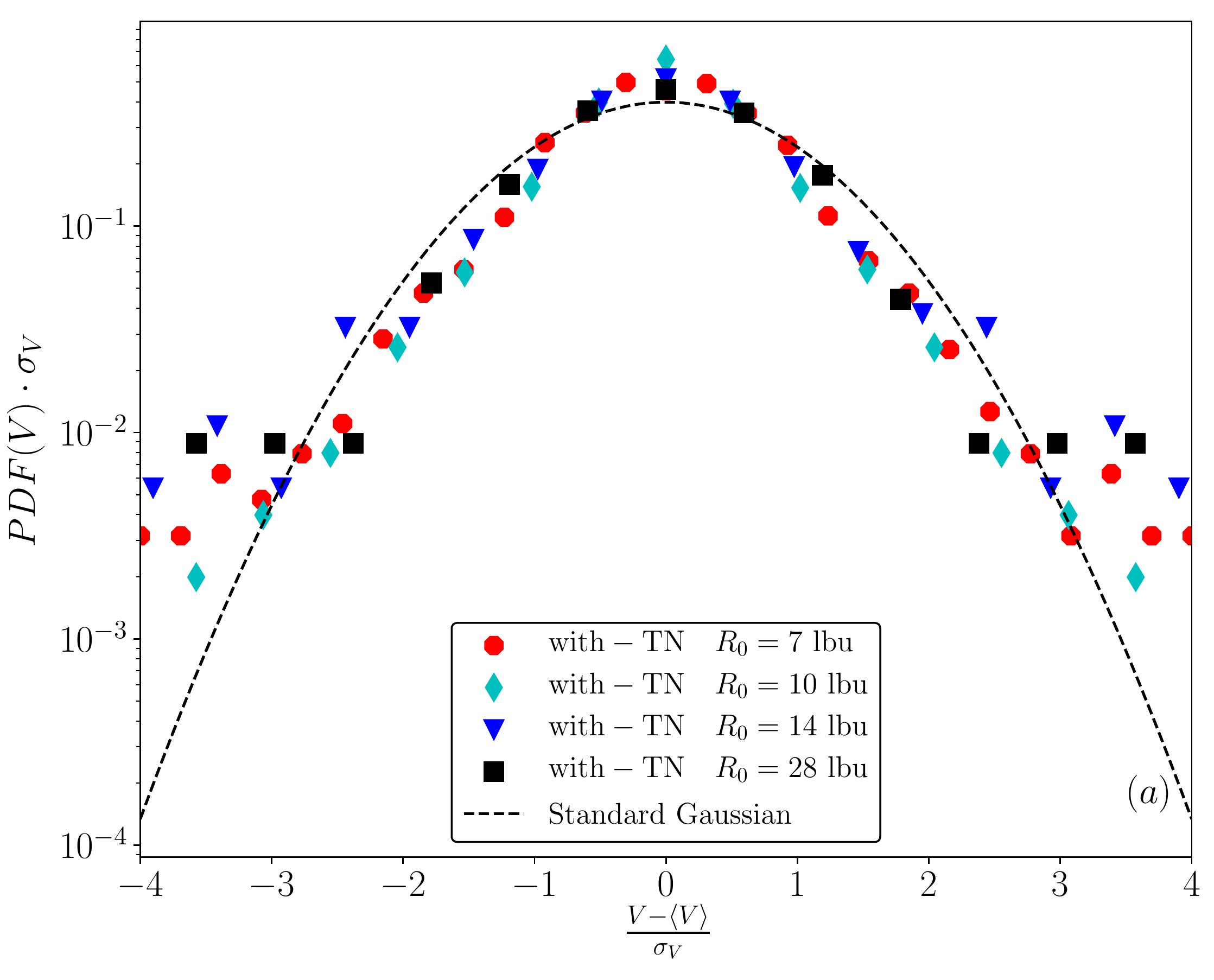}\label{fig:normMajorSata}}
\subfigure{\includegraphics[height = 0.38\linewidth]{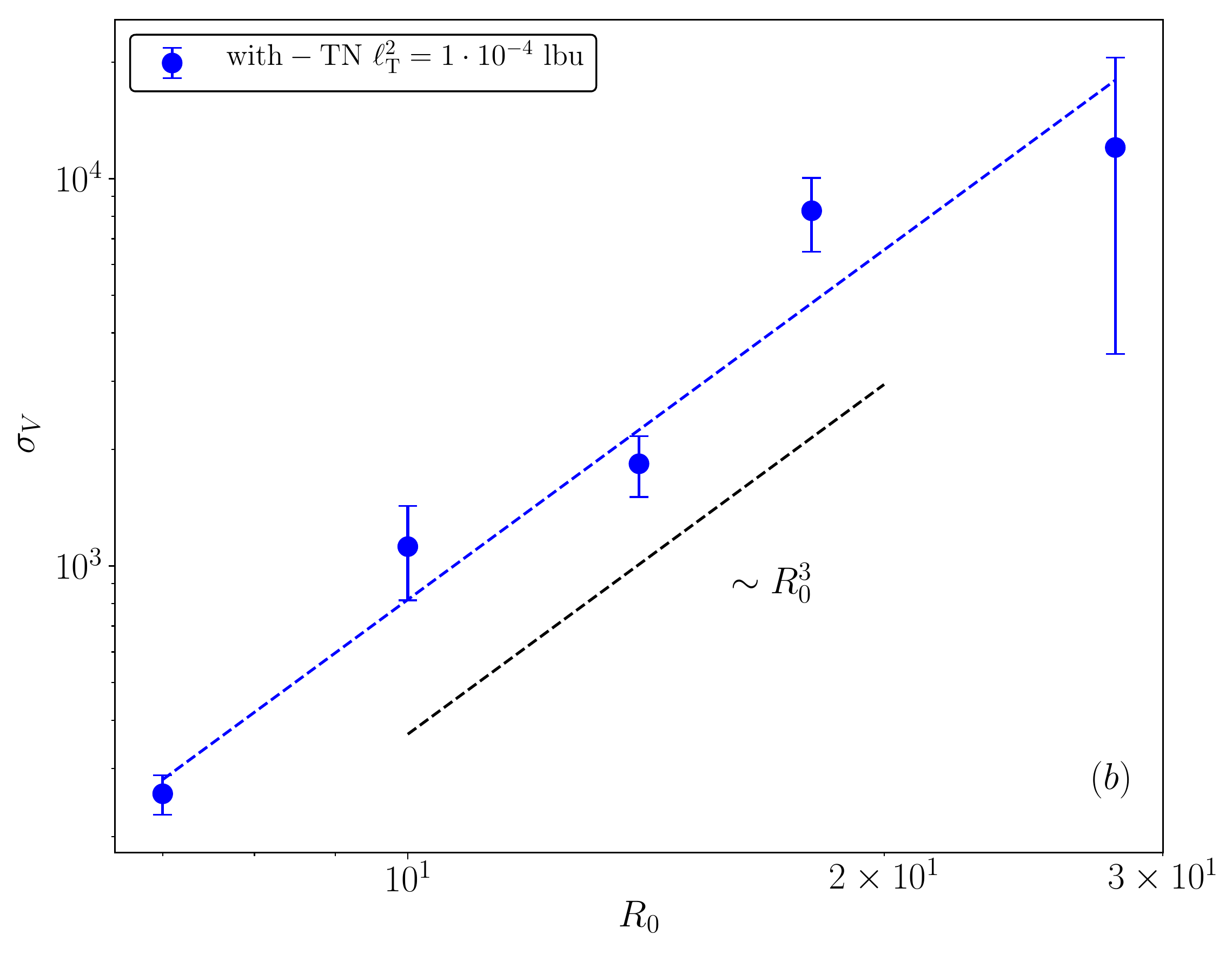}\label{fig:normMajorSatb}}
\end{minipage}
\caption{Panel (a): standardized PDFs of the droplet volumes at changing the ligament radius $R_0$ for fixed thermal length $\ellT$. Only data for the ``with-TN'' protocol are analyzed. Panel (b): standard deviations $\sigma_V$ at changing ligament radius $R_0$, for fixed thermal length squared $\ellT^2 = 1\cdot 10^{-4}$. Error bars are estimated from the standard deviation.}
\label{fig:PDFVolumesb}
\end{figure}

In this section we study the droplet volumes after break-up. For this analysis, the droplet volumes are measured using the marching tetrahedra method~\cite{doi1991efficient} with Paraview software and a Python interface. This ensures a very accurate measurement of the droplets volume which is a key ingredient to differentiate the small changes induced by thermal fluctuations. In~\cref{fig:volumeCompare} we show the PDFs for the droplets volumes for the ``with-TN'' protocol at changing the thermal length $\ellT$ for fixed ligament radius $R_0$ (top row, Panels (a)-(c)) or at changing the ligament radius $R_0$ for fixed thermal length (bottom row, Panels (d)-(f)). We show only the volume distribution  of the ``mother'' droplets (see \cref{fig:ligament}): this is done to limit the range of $V$ and allow for a more insightful comparison. We have separately analyzed the PDFs of the satellite droplets and the conclusions drawn for \cref{fig:volumeCompare} are valid for them as well. From  \cref{fig:volumeCompare}, we can see that when the thermal length increases the PDFs develop larger standard deviation (Panels (a)-(c)), the same haFppens when we increase the ligament radius $R_0$ at fixed thermal length (Panels (d)-(f)). The shape of the PDFs and the dependency of the standard deviation $\sigma_V$ on both $\ellT^2$ and $R_0$ are quantitatively summarized in \cref{fig:normalPdf} and \cref{fig:normMajorSata}. In particular, in \cref{fig:normalPdf} we report the standardized PDFs at changing the thermal lengths for fixed ligament radius for both protocols. We observe that the rescaled PDFs collapse well on the same master curve, independently of the thermal length and the simulation protocol. This indicates that the presence of randomness in the initial condition plays the major role in determining the shape of the distributions. A tendency towards a slight sub-Gaussain behaviour is detectable for small (normalized) volume fluctuations, whereas larger fluctuations are associated with tails higher than Gaussian. The analysis of the standard deviation $\sigma_V$ (\cref{fig:STDlaw}) shows that in the ``with-TN'' protocol the droplets polydispersity is enhanced by a factors around 40\% with respect to what we obtain in the ``without-TN'' protocol. In both cases, however, signatures of a scaling law with exponent $\sim 0.14$ are obtained by fitting simulations data. Even though we are not able to provide analytical explanation for the scaling law, however, we confirm the observed scaling law by comparing with the sharp interface hydrodynamics in the following \cref{sec:hydroLBM}. In \cref{fig:PDFVolumesb} we show the results of  a similar analysis but at changing the ligament radius for fixed thermal length for the physically relevant ``with-TN'' protocol. Again, we observe that the rescaled PDFs collapse well on the same master curve. Data are more scattered with respect to \cref{fig:PDFVolumes} due to the smaller number of simulations used to compute the PDFs. Overall, we notice that the PDFs reported in \cref{fig:normalPdf} and \cref{fig:normMajorSata} display fatter tails with respect to a Gaussian distribution. If from one side one could say that the statistics accumulated on the tails may be not enough to precisely quantify them, from the other side we will also present in \cref{sec:hydroLBM} data on sharp interface hydrodynamics supporting the view that those tails are definitively non Gaussian. Regarding the standard deviation (\cref{fig:normMajorSatb}), we observe a scaling law close to 3, $\sigma_{V} \sim R_{0}^{3}$ that is what one would expect based on pure geometrical considerations. Regarding the connection between the distribution of volumes and break-up times, one could say that if the former is Gaussian and the latter  log-normal, then an Arrhenius-like scenario could be invoked to connect the two, i.e.  $t^{*} \sim e^{\Delta E/k_{B}T}$, with $\Delta E$ some energy contribution proportional to the volume variation. However, it must be noted that the volume distribution in \cref{fig:normalPdf} and \cref{fig:normMajorSata} are slightly non-Gaussian, hence the above relation would imply also some departure from log-normality for the time distribution, further data would be needed  to clarify this issue.


\subsection{Comparison with sharp-interface hydrodynamics}\label{sec:hydroLBM}
\begin{figure}[H]
\begin{center}
\begin{minipage}{1.0\textwidth}
\subfigure{\includegraphics[height = 0.38\linewidth] {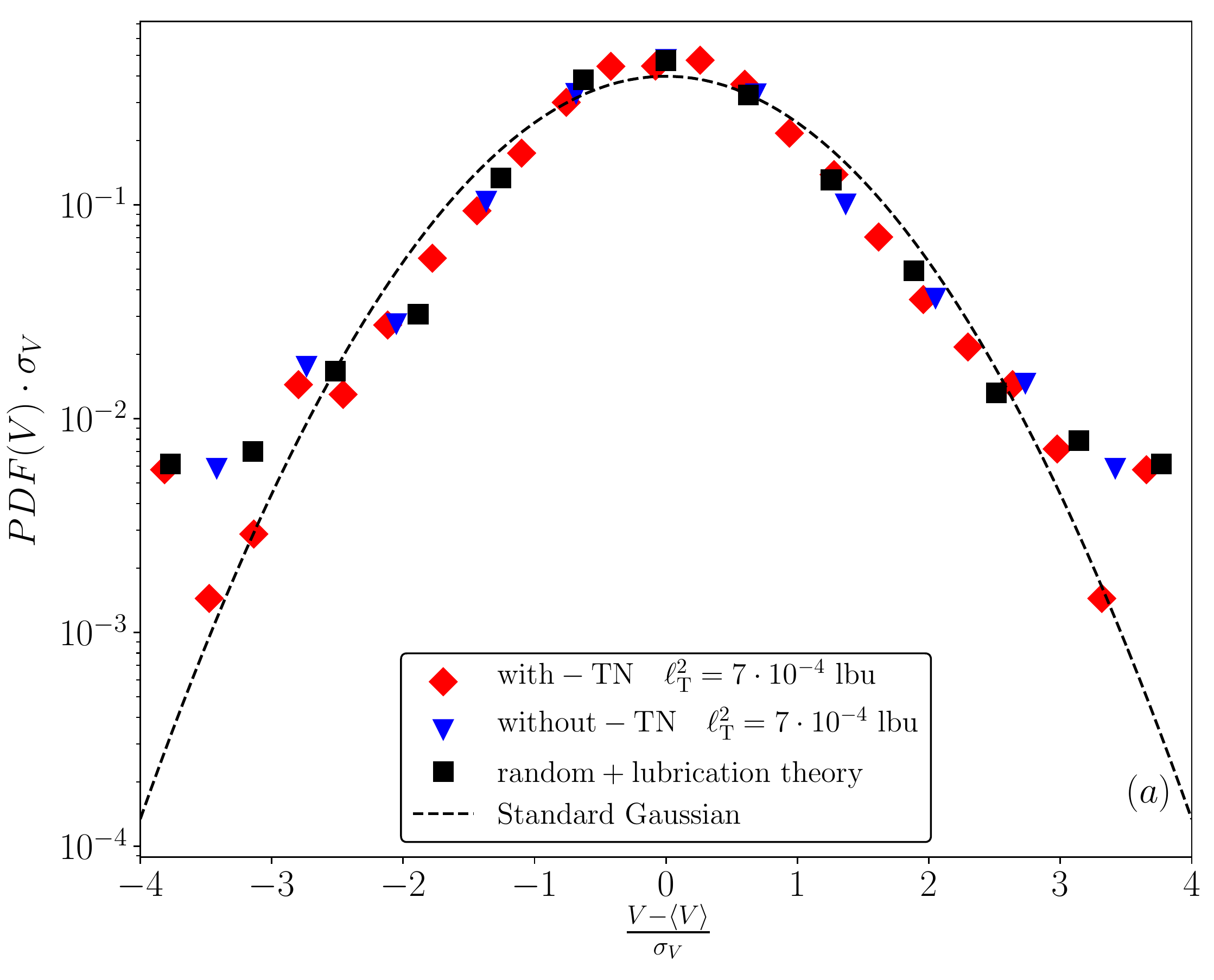}\label{fig:normLub}}
\subfigure{\includegraphics[height = 0.38\linewidth]{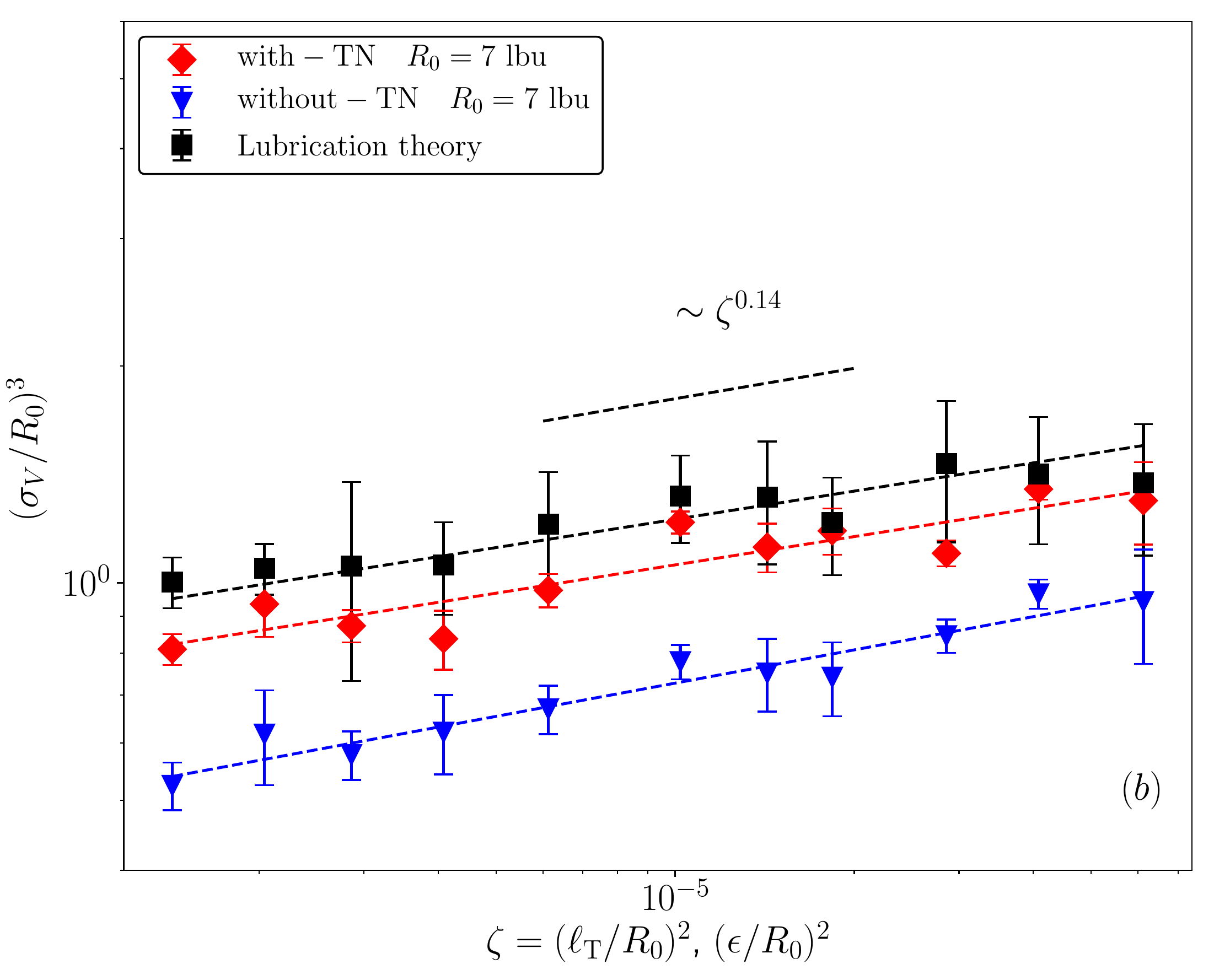}\label{fig:stdFlucVsNonFluc(b)}}
\end{minipage}
\end{center}
\caption{Panel (a): comparison among the standardized PDFs of the droplet volumes following the two LB evolution protocols or the lubrication equations with initial random disturbance (see text for details). Panel (b): rescaled standard deviation $(\sigma_V/R_{0})^{3}$ at changing the rescaled thermal length squared $(\ellT/R_{0})^2$ and rescaled Gaussian noise amplitude $(\epsilon/R_{0})^2$ for lubrication theory. Error bars are estimated from the standard deviation of different groups of the configurations.}
\label{fig:lubrication}
\end{figure}

In the previous sections we observed that rescaled PDFs of droplet volumes display a sub-Gaussian shape for small volume fluctuations, and higher tails for larger fluctuations (see \cref{fig:normalPdf} and \cref{fig:normMajorSata}); moreover the standard deviation displays signatures of scaling laws in $\ellT^2$ (see \cref{fig:STDlaw}). To better reveal the origin of the shape of the PDFs and the scaling law for the standard deviation, we conducted additional numerical simulations with deterministic sharp interface hydrodynamics~\cite{eggers1993universal, eggers1994drop}. This comparison with sharp interface hydrodynamics can also elucidate the role of the diffuse interfaces which are inherent to the LB approach. We numerically considered an axisymmetric formulation of the lubrication equation of sharp interface hydrodynamics~\cite{eggers1993universal,eggers1994drop,driessen2011regularised} using a finite difference scheme with total variation diminishing method~\cite{driessen2011regularised}. In this approach, the periodic axisymmetric ligament is placed along the $x$ axis and the whole evolution is described by its height, $h(x,t)$, and its axial velocity, with $v(x,t)$. The dimensionless lubrication equation becomes: 
\be
\partial_{t} h^{2}+ \partial_x (h^{2}v) = 0
\ee
\begin{equation}
\partial_{t} v+ v \partial_x v= -\partial_x P_{\st{lap}} + 3 \mbox{Oh}\, h^{-2}\, \partial_x(h^{2}\partial_x v)
\end{equation}
where $P_{\st{lap}}$ is the Laplace pressure that can be written as
\begin{equation}
P_{\st{lap}}=\left [ \frac{1}{h(1+(\partial_x h)^{2})^{\frac{1}{2}}} - \frac{\partial_{xx} h}{(1+(\partial_x{h})^{2})^{\frac{3}{2}}}\right ].
\end{equation}
To study the droplet size distributions in the sharp interface hydrodynamics approach, we imposed an initial perturbation on the radius in the form $R_0+\epsilon \Xi(x)$, with $\Xi(x)$ a random Gaussian variable with unitary variance and zero mean and $\epsilon$ a small number that is the analogous of $\ellT$ that we have used in the LB simulations. By varying the realization of the variable $\Xi(x)$ in the initial conditions we can compute the PDFs for droplet volumes. In other words, the hydrodynamic solver is the sharp interface counterpart of the ``without-TN'' protocol. The ensemble that we consider now is made of 4000 simulations, which is larger than what we used for the LB simulations (see \cref{tab:parameter_space}). This will allow us to see how much of the observed behaviour for the LB simulations can depend on the statistics analyzed. Results are displayed in \cref{fig:lubrication}. We notice that we obtain fatter tails with respect to a Gaussian distribution. Moreover, the agreement of the rescaled PDFs is remarkable (Panel (a)), suggesting  that the main effect fixing the shape of the standardized PDF is due to the destabilization of multiple modes in the deterministic hydrodynamic framework.  Further insight is conveyed by the analysis of the standard deviation $\sigma_V$ (Panel (b)). Notice that both $\sigma_V$, $\epsilon$, $\ellT$ have been made dimensionless with the characteristic scale $R_0$ in order to allow for a fare comparison. The direct comparison against the ``with-TN'' protocol reveals that the two curves are offset by a constant factor. This may be due to the diffusive interface of the LB or the different viscosity ratio between the LB simulations (where the dynamic viscosity ratio is 1) and the sharp interface approach (where the dynamic viscosity ratio is infinity). Further investigation needs to be performed to clarify these points. Nevertheless, we wish to point out that the scaling properties with respect to the characteristic amplitude of the initial perturbation is consistent in all three cases.

\section{Conclusions}\label{sec:conclusions}
We have used numerical simulations based on fluctuating multicomponent lattice Boltzmann (LB) models~\cite{Belardinelli15} to study the effects of thermal fluctuations on the break-up time of a liquid ligament and the associated polydispersity in droplets volumes after break-up. To quantitatively understand the role of thermal fluctuations during the dynamical process of the break-up, we have designed two different simulation protocols that allowed to evolve {\it with} or {\it without} thermal fluctuations a random initial condition realized over the ligament interface. From one side the thermal fluctuations allow to speed-up the break-up process~\cite{Hennequin06, Eggers02} and to obtain larger polydispersity; from the other side the shape of the resulting PDF for droplet volume appears to be largely generated by a dynamical process that does not involve fluctuating hydrodynamics. The leading mechanism is that of a fastest-growing mode that is destabilized by the Plateau\texttt{-}Rayleigh instability, and other unstable modes (growing at smaller rate) that provide -- if initialized with random phases and amplitudes -- an effective noise broadening the final distributions of droplet volumes. As a future perspective, there are various interesting issues to be investigated. For example, it could be an interesting challenging problem to predict the observed shape for the PDFs directly from sharp interface hydrodynamics, as well as the scaling laws for the standard deviations or the break-up time. We also remark that the thermal lengths that we explored are quite small in comparison to the ligament radius. Hence, it could be a challenging computational task to extend our study in a range of parameters with larger thermal lengths~\cite{Hennequin06,Petit12}.
\section{ACKNOWLEDGEMENTS}
The authors would like to kindly acknowledge funding from the European Union's Horizon 2020 research and innovation programme under the Marie Sk\l{}odowska-Curie grant agreement No 642069(European Joint Doctorate Programme ``HPC-LEAP"). L. Biferale and M. Sbragaglia acknowledge the support from the ERC Grant No 339032. We acknowledge fruitful discussions and exchanges with J. Eggers, M. Sega and D. Belardinelli. We also acknowledge J.L.L. Lopez for contributions in the early stage of the work.

\bibliographystyle{spphys}
\bibliography{prex}
\end{document}